\documentclass[12pt,preprint]{aastex}

\slugcomment{submitted to the Astrophysical Journal}
\shorttitle{Hardness Ratios in Poisson Regime}
\shortauthors{T.~Park et al.}

\begin{document}

\newcommand\up{\raisebox{1.7ex}[0pt]}
\newcommand\divi{\multicolumn{1}{c|}}
\newcommand\mdivi{\multicolumn{2}{c|}}
\newcommand\multi{\multicolumn{2}{c}}
\newcommand\ind{\stackrel{\rm indep}{\sim}}
\newcommand\iid{\stackrel{\rm iid}{\sim}}
\newcommand\define{\stackrel{\triangle}{=}}
\renewcommand\a{_{\rm aug}}
\renewcommand\|{\,|\,}

\newcommand\E{{\rm E}}
\newcommand\Var{{\rm Var}}
\newcommand\var{{\rm var}}
\newcommand\N{{\rm N}}
\newcommand\TN{{\rm Truncated \:\: Normal}}
\newcommand\MVN{{\rm MVN}}
\newcommand\Inv{{\rm Inv}}
\newcommand\IG{{\rm IG}}
\newcommand\Wi{{\rm Wishart}}
\newcommand\Gam{{\rm Gamma}}
\newcommand\Be{{\rm Beta}}
\newcommand\Binom{{\rm Binomial}}
\newcommand\Pois{{\rm Poisson}}
\newcommand\Bin{{\rm Binomial}}
\newcommand\Ber{{\rm Bernoulli}}
\newcommand\Multin{{\rm Multinomial}}
\newcommand\Cov{{\rm Cov}}
\newcommand\argmax{{\rm argmax}}
\newcommand\al{\alpha}
\newcommand\be{\beta}
\newcommand\sig{\sigma}
\newcommand\Sig{\Sigma}
\renewcommand\th{\theta}
\newcommand\Th{\Theta}
\newcommand\eps{\varepsilon}
\newcommand{\balpha}{\mbox{\boldmath{$\alpha$}}}
\newcommand\Ymis{Y_{\rm mis}}
\newcommand\Yobs{Y_{\rm obs}}
\newcommand\Yaug{Y_{\rm aug}}
\newcommand\DM{{\rm DM}}
\newcommand\EM{{\rm EM}}

\newcommand\bkg{_{\cal bkg}}
\newcommand\R{{\cal R}}
\newcommand\HR{{\cal HR}}
\newcommand\C{{\rm C}}
\newcommand\Bmis{_{\rm mis}}
\newcommand\Bobs{_{\rm obs}}
\newcommand\Baug{_{\rm aug}}
\newcommand\D{{\cal D}}
\newcommand\Da{{\cal D}_\alpha}
\newcommand\Daa{{\cal D}_{\alpha^\star,1}}
\newcommand\Daone{{\cal D}_{\alpha,1}}
\newcommand\T{{\cal T}}
\newcommand\K{{\cal K}}
\newcommand\M{{\cal M}}
\newcommand\I{{\rm I}}
\newcommand\J{{\cal J}}
\newcommand\Q{{\cal Q}}
\renewcommand\L{{\cal L}}
\newcommand\om{\omega}
\newcommand\vrho{\varrho}
\newcommand\itoN{^N_{i=1}}
\newcommand\ptoP{^P_{p=1}}
\newcommand\iton{^n_{i=1}}
\newcommand\jtoJ{^J_{j=1}}
\newcommand\ktoK{^K_{k=1}}
\newcommand\inv{^{-1}}
\newcommand\ij{_{ij}}
\newcommand\twoth{_{\btwo,\th}}
\newcommand\pr{^{\prime}}
\renewcommand\*{^{\star}}
\newcommand\wt{\widetilde}
\newcommand\ds{\displaystyle}

\newcommand\kase{{C{\scriptsize ASE}}}
\newcommand\kases{C{\scriptsize ASES}}
\newcommand\step{S{\scriptsize TEP}}
\newcommand\steps{S{\scriptsize TEPS}}
\newcommand\mcmc{{\sc mcmc}}
\newcommand\chandra{{\sl Chandra}}

\def\X{{\bf X}}
\def\B{{\bf B}}
\def\Y{{\bf Y}}
\def\Ya{\dot {\bf Y}}
\def\Yb{\ddot {\bf Y}}
\newcommand\Yc{{\bf Y}\kern -.65pc{}^{^{\textstyle\cdot\kern -3pt\cdot\kern -3pt\cdot}}
          \hskip -0.75pc\phantom{Y}}

\def\updot{^{\textstyle\cdot}}
\def\obs{^{\rm obs}}

\def\lam{\lambda}
\def\Lam{\Lambda}
\def\*{\star}
\def\l{\left}
\def\r{\right}

\def\eff{{e}}


\title{\bf Bayesian Estimation of Hardness Ratios: Modeling and Computations}

\author{Taeyoung~Park\altaffilmark{1},
Vinay~L.~Kashyap\altaffilmark{2},
Aneta~Siemiginowska\altaffilmark{2},
David~A.~van~Dyk\altaffilmark{3},
Andreas~Zezas\altaffilmark{2},
Craig~Heinke\altaffilmark{4},
\& Bradford~J.~Wargelin\altaffilmark{2}
}

\affil{$^1$Department of Statistics\\
Harvard University \\
One Oxford Street, Cambridge, MA 02138
\email{tpark@stat.harvard.edu}}

\affil{$^2$Smithsonian Astrophysical Observatory, \\
60 Garden Street, Cambridge, MA 02138
\email{kashyap@head.cfa.harvard.edu \\
aneta@head.cfa.harvard.edu \\
azezas@head.cfa.harvard.edu \\
bwargelin@head.cfa.harvard.edu}}

\affil{$^3$Department of Statistics\\
University of California, \\
364 ICS Bldg One, Irvine, CA 92697-1250
\email{dvd@ics.uci.edu}}

\affil{$^4$Northwestern University,\\
2131 Tech Drive, Evanston, IL 60208
\email{cheinke@northwestern.edu}}

\begin{abstract}

A commonly used measure to summarize the nature of a photon
spectrum is the so-called Hardness Ratio, which compares the
number of counts observed in different passbands.  The
hardness ratio is especially useful to distinguish between
and categorize weak sources as a proxy for detailed spectral
fitting.  However, in this regime classical methods of error
propagation fail, and the estimates of spectral hardness
become unreliable.  Here we develop a rigorous statistical
treatment of hardness ratios that properly deals with
detected photons as independent Poisson random variables and
correctly deals with the non-Gaussian nature of the error
propagation.  The method is Bayesian in nature, and thus can
be generalized to carry out a multitude of source-population--based
analyses.  We verify our method with simulation studies, and
compare it with the classical method.  We apply this method
to real world examples, such as the identification of candidate
quiescent Low-mass X-ray binaries in globular clusters, and
tracking the time evolution of a flare on a low-mass star.

\end{abstract}




\section{Introduction}
\label{park:sec:intro}

The shape of the spectrum of an astronomical source is highly
informative as to the physical processes at the source.  But
often detailed spectral fitting is not feasible due to various
constraints, such as the need to analyze a large and homogeneous
sample for population studies, or there being insufficient counts
to carry out a useful fit, etc.  In such cases, a hardness ratio,
which requires the measurement of accumulated counts in two or more broad
passbands, becomes a useful measure to quantify and characterize
the source spectrum.\footnote{
In the context of \chandra\ data,
the passbands often used are called {\sl soft} (0.3-0.9~keV),
{\sl medium} (0.9-2.5~keV), and {\sl hard} (2.5-8.0~keV) bands.
Sometimes the {\sl soft} and {\sl medium} bands are merged into
a single band, also called the {\sl soft} band (e.g., 0.5-2~keV).
In all cases, note that the energies refer to the nominal detector
PHA or PI channel boundaries and not to the intrinsic photon energies.
The choice of energy or PI channel range is flexible and situational.
}
A hardness ratio is defined as either the ratio of the counts in two bands
called the {\sl soft} and {\sl hard} bands, or a monotonic function of
this ratio.  We consider three types of hardness ratios,
\begin{eqnarray}
\nonumber {\rm Simple~Ratio,}~~&~\R \equiv& \frac{S}{H} \\
\nonumber {\rm Color,}~~&~\C \equiv& \log_{10}\left(\frac{S}{H}\right) \\
{\rm Fractional~Difference,}~~&\HR \equiv& \frac{H-S}{H+S}
\label{e:RCHR}
\end{eqnarray}
where $S$ and $H$ are the source counts in the two bands, called the
{\sl soft} and {\sl hard} passbands.  The simple formulae above are
modified for the existence of background counts and instrumental
effective areas.

Spectral colors in optical astronomy, defined by the standard
optical filters (e.g., UBVRIJK, U-B, R-I, etc), are well known
and constitute the main observables in astronomy.  They have
been widely applied to characterize populations of sources and their
evolution.
The hardness ratio concept was adopted in X-ray astronomy
for early X-ray detectors (mainly proportional counter detectors)
which had only a limited energy resolution. The first application of
the X-ray hardness ratio was presented in the X-ray variability
studies with SAS-3 observations by Bradt et al.\ (1976). Zamorani et
al.\ (1981) investigated the X-ray hardness ratio for a sample of
quasars observed with the {\it Einstein} X-ray Observatory. They
calculated each source intensity in two energy bands (1.2-3~keV and
0.5-1.2~keV) and discussed a possible evolution of the hardness ratio
(the ratio of both intensities) with redshift for their X-ray sample
of 27 quasars. Similar ratios have been used in surveys of stars
(Vaiana et al.\ 1981), galaxies (e.g., Kim, Fabbiano, \& Trinchieri 1992) and X-ray binaries (Tuohy et al.\ 1978) studied with the {\it
Einstein} and earlier observatories.  In the case of X-ray binaries in
particular, they were used to define two different classes of sources
(Z-track and atoll sources; Hasinger \& van der Klis, 1989) depending on
their time evolution on $\HR$ v/s $\HR$ diagrams.  Since then the concept of X-ray hardness ratio
has been developed and broadly applied in a number of cases, most
recently in the analysis of large data samples from the \chandra\
X-ray Observatory (Weisskopf et al.\ 2000) and XMM-{\sl Newton}
(Jansen et al.\ 2001).

Advanced X-ray missions such as \chandra\ and XMM-{\sl Newton}
allow for the detection of many very faint sources in deep
X-ray surveys (see review by Brandt \& Hasinger 2005).
For example, in a typical observation of a galaxy, several tens and in
some cases even hundreds of sources are detected, most of which have
fewer than 50 counts, and many have only a few counts.
Similar types of sources are detected in the
ChaMP serendipitous survey (Kim et al.\ 2005).  
They provide the best quality data to date for studying
source populations and their evolution. The most interesting sources
are the sources with the smallest number of detected counts, because
they have never been observed before.
Further, the number of faint sources increases with improved
sensitivity limits, i.e., there are more faint sources in
deeper observations.
Because these faint sources have only a few counts, hardness ratios are
commonly used to study properties of their X-ray emission and absorption
(e.g., Alexander et al.\ 2001, Brandt et al.\ 2001a,
Brandt et al.\ 2001b, Giacconi et al.\ 2001, Silverman et al.\ 2005).
With long observations the
background counts increase, so the background contribution becomes
significant and background subtraction fails to work. Still
the background contribution must be taken into account to evaluate
the source counts.

The sensitivity of X-ray telescopes is usually energy dependent, and
thus the classical hardness ratio method can only be applied to
observations made with the same instrument. Usually the measurements
carry both detection errors and background contamination. In a typical
hardness ratio calculation the background is subtracted from the data
and only net counts are used. This background subtraction is not a
good solution, especially in the low counts regime (see van Dyk et al.\
2001).
In general the Gaussian assumption present in the classical method
(see \S\S~\ref{park:sec:classic}~and~\ref{park:sec:verify}) is not
appropriate for faint sources in the presence of a significant background.
Therefore, the classical approach to calculating the hardness ratio
and the errors is inappropriate for low counts.  Instead, adopting a
Poisson distribution as we do here (see below), hardness ratios
can be reliably computed for both low and high counts cases.
The Bayesian approach allows us to include the
information about the background, difference in collecting area,
effective areas and exposure times between the source and background
regions.

Our Bayesian model-based approach follows a pattern of ever more
sophisticated statistical methods that are being developed and applied
to solve outstanding quantitative challenges in empirical Astronomy
and Astrophysics. Bayesian model-based methods for high-energy
high-resolution spectral analysis can be found for example in
Kashyap \& Drake (1998),
van Dyk et al. (2001), van Dyk and Hans (2002), Protassov et al.\ (2002),
van Dyk and Kang (2004), Gillessen \& Harney (2005), and
Park et al.\ (2006, in preparation). More generally, the
monographs on Statistical Challenges in Modern Astronomy (Feigelson and
Babu, 1992, 2003, Babu and Feigelson, 1997) and the special issue of {\it
Statistical Science} devoted to Astronomy (May 2004) illustrate the
wide breadth of statistical methods applied to an equal diversity of
problems in astronomy and astrophysics.

Here, we discuss the classical method and present the fully Bayesian
method for calculating the hardness ratio for counts data.\footnote{
A Fortran and C-based program which calculates the hardness ratios
following the methods described in this paper is available for download
from
\url{{\tt http://hea-www.harvard.edu/AstroStat/BEHR/}}
}

\subsection{The Classical Method}
\label{park:sec:classic}

A conventional hardness ratio is calculated as set out in
Equations~\ref{e:RCHR}, where $S$ and $H$ are the ``soft'' and ``hard''
counts accumulated in two non-overlapping passbands.
In general, an observation will be contaminated by background counts,
and this must be accounted for in the estimate of the hardness ratios.
The background is usually estimated from an annular region surrounding
the source of interest, or from a suitable representative region on
the detector that is reliably devoid of sources.
The difference in the exposure time and aperture area of
source and background observations are summarized by a known constant
$r$ for which the expected background counts in the source exposure
area are adjusted. With the background counts in the soft band ($B_S$)
and the hard band ($B_H$) collected in an area of $r$ times the source
region, the hardness ratio is generalized to
\begin{eqnarray}
\nonumber \R &=& \frac{S-B_S/{r}}{H-B_H/{r}} \\
\nonumber \C &=& \log_{10}\left(\frac{S-B_S/{r}}{H-B_H/{r}}\right) \\
\HR &=& \frac{(H-B_H/{r})-(S-B_S/{r})}{(H-B_H/{r})+(S-B_S/{r})}
\label{e:bgRCHR}
\end{eqnarray}
The adjusted counts in the background exposure area are
directly subtracted from those in the source exposure area.
The above equations can be further modified to account for variations
of the detector effective areas by including them in the constant $r$,
in which case the constants for the two bands will be
different.\footnote{
Gross differences in detector sensitivity,
e.g., between different observations, or for sources at different
off-axis angles, can be accounted for with additional multiplicative
factors modifying the background-subtracted counts.  This is however
not statistically meaningful and can lead to unreliable estimates
and error bars in the small counts regime.  Instead, we apply such
scaling factors directly to the source intensities (see
Equation~\ref{park:eq:SH}).
}
Errors are propagated assuming a Gaussian regime, i.e.,
\begin{eqnarray}
\nonumber
 \sig_{{\R}} &=& \ds\frac{S-B_S/{r}}{H-B_H/{r}}
	\sqrt{\:\frac{\sig_{S}^2+\sig_{B_S}^2/{r}^2}{(S-B_S/r)^2} 
	+ \frac{\sig_{H}^2+\sig_{B_H}^2/{r}^2}{(H-B_H/r)^2}} \\
\nonumber
 \sig_{{\C}} &=& \ds\frac{1}{\ln(10)} \ 
	\sqrt{\:\frac{\sig_{S}^2+\sig_{B_S}^2/{r}^2}{(S-B_S/r)^2} 
	+ \frac{\sig_{H}^2+\sig_{B_H}^2/{r}^2}{(H-B_H/r)^2}} \\
 \sig_{{\HR}} &=& \ds\frac{2\: \sqrt{(H-B_H/r)^2\big(\sig_{S}^2
	+\sig_{B_S}^2/{r}^2\big) + (S-B_S/r)^2\big(\sig_{H}^2
	+\sig_{B_H}^2/{r}^2\big)}}{\big[(H - B_H/r)+(S - B_S/r)\big]^2}
\label{e:sigbgRCHR}
\end{eqnarray}
where $\sig_{X_S}$, $\sig_{X_H}$, $\sig_{B_S}$, and $\sig_{B_H}$ are
typically approximated with the Gehrels prescription (Gehrels 1986)
\begin{equation}
\sig_{X} \approx \sqrt{X+0.75}+1 \,,
\label{park:eq:sig2}
\end{equation}
where $X$ are the observed counts, and the deviation from $\sqrt{X}$
is due to identifying the 68\% Gaussian (1$\sigma$) deviation with the
$16^{\rm th}-84^{\rm th}$ percentile range of the Poisson distribution.
In addition to the approximation of a Poisson with a faux Gaussian
distribution, classical error-propagation also fails on a fundamental
level to properly describe the variance in the hardness ratios.
Because $\R$ is strictly positive, its probability distribution is
skewed and its width cannot be described simply by $\sig_{{\R}}$.
The fractional difference $\HR$ is better behaved, but the
range limits of $[-1,+1]$ cause the coverage rates (see
\S\ref{park:sec:verify}) to be incorrect.  The color $\C$ is the
best behaved, but the error range will be asymmetrical when the errors
are large.  Moreover, it has been demonstrated (van Dyk et al.\ 2001)
that using background subtracted source counts (which is the case in
the method outlined above) gives a biased estimate of the source
intensity, especially for very faint sources.
Furthermore, although we can account for gross differences in the
detector sensitivity between observations, this is not done in a
statistically correct manner, but instead simply by rescaling the
background-subtracted counts in each band using additional
pre-computed correction factors.
Finally, the classical hardness ratio method cannot be applied
when a source is not detected in one of the two bands (which is
quite common since CCD detectors are more sensitive in the soft band).

An alternative method  based on the median (or some quantile) of the
distribution of the energies of the detected counts in a source was
suggested by Hong et al.\ (2004).  This method can be very
powerful for classifying faint sources, but is not suited for
comparisons of spectral properties between observations with
different instruments.  
In this method, because the spectral parameter grids are highly
non-linear and multi-valued in quantile-quantile space, it is also
necessary that the spectral shape of the source being analyzed be
known in order to interpret the numerical values.

We have therefore developed fully Bayesian approaches to appropriately
compute hardness ratios and their errors, by properly accounting
for the Poissonian nature of the observations.  The value of our method
is heightened in several astrophysically meaningful cases.
In \S\ref{park:sec:model}, we describe the model structure we
adopt.  In \S\ref{park:sec:verify}, we carry out simulations to
compare our method with the classical method.  In \S\ref{park:sec:apply},
we outline various applications of our method; we demonstrate its
use in identifying candidate quiescent Low-mass X-ray binary 
sources in a globular cluster and
in studying the evolution of a stellar X-ray flare.
In \S\ref{park:sec:discuss}, we discuss some nuances of our method such
as the adopted prior distributions, as well as the advantages and
limitations.  A summary of the method and its advantages
is presented in \S\ref{park:sec:conclusion}.
A detailed mathematical derivation of the posterior probability
distributions of the hardness ratios and the computational methods
we use to calculate them are described in Appendix~\ref{sec:appA},
and a detailed comparison of the effects of priors is described in
Appendix~\ref{sec:appB}.

\section{Modeling the Hardness Ratios}
\label{park:sec:model}

\subsection{The Poisson Case}
\label{park:sec:poisson}

Hardness ratios are often used to describe faint sources with very few
counts; it is not uncommon for either or both of the soft and hard
counts to be small enough that the source is undetected in that
band. The classical method (\S\ref{park:sec:classic}) generally fails
to produce reliable estimates (or upper and lower limits) in such cases, and indeed, it is
sometimes not even possible to apply it, such as when the background
is large and simple subtraction results in a negative estimate for the
source counts.  Thus, instead of the Gaussian assumptions on the
detected counts, we directly model them as an inhomogeneous Poisson
process.

Since the observed
counts are the sum of counts due to the source and the background,
and because the sum of two independent Poisson random variables is a
Poisson random variable with the sum of two Poisson intensities,
we may write\footnote{
Note that throughout
this paper, Greek letters indicate unobserved quantities and
Roman letters denote observed data values (except for $\R$, $\C$,
and $\HR$, which are used to represent the hardness ratio model 
parameters in the Poisson case).}
\begin{eqnarray}
\:S \sim \Pois(\eff_S \cdot (\lam_S+\xi_S)) &\mbox{and}& 
H \sim \Pois(\eff_H \cdot (\lam_H+\xi_H)), \label{park:eq:SH}
\end{eqnarray}
where $S$ and $H$ are independent data points, $\lam_S$ and $\lam_H$ are the expected
soft and hard source counts intensities, $\xi_S$ and $\xi_H$ are the expected soft
and hard background counts intensities, and $\eff_S$ and $\eff_H$ are correction
factors that take into account variations in effective areas,
exposure times, and other instrumental effects.\footnote{
Note
that in general the effective area changes within an adopted passband,
so the chosen values of the correction factor are averages that
are calculated after making some prior assumptions about the
underlying source spectrum.  Their values may change depending on
the analysis results.
}
The observed background counts 
are also modeled as independent Poisson random variables,
\begin{eqnarray}
  B_S \sim \Pois({r}\cdot \eff_S \cdot \xi_S) &\mbox{and}& 
  B_H \sim \Pois({r}\cdot \eff_H \cdot \xi_H)  \label{park:eq:B}
\end{eqnarray}
where $\xi$ is scaled by a known correction factor $r$,
to account for differences in source and background areas and
sensitivities.
We implicitly assume throughout that the observed data are
discrete photon events, or counts.  The intensities $\lambda$
and $\xi$ may have other units, determined by how the
instrument efficiency is supplied.  Thus, if $\eff_S$ is given
in [ct~s~cm$^{2}$~ph$^{-1}$], then $\lam_S$ will have units
[ph~s$^{-1}$~cm$^{-2}$].  Alternately, $\eff_S$ could 
describe the exposure time in [s], in which case $\lam_S$
will have units of [ct s$^{-1}$].  In the case where
$\lam_S$ and $\lam_H$ themselves have the units [ct], then
$\eff_S$ and $\eff_H$ are dimensionless, and usually taken
to be unity.  This allows us to compute
hardness ratios of fluxes and count rates as well as counts.
We place no restriction on the units of the model intensities
(other than that $\lambda$ and $\xi$ have the same units);
they are entirely determined by the problem at hand. Thus,
our analysis is valid for any instrumental configuration or
combination, and the only requirement is that the observed
data be described by a Poisson distribution.

Given the expected source counts intensities ($\lam_S$ and $\lam_H$),
it is legitimate for the hardness ratio to be rewritten as
\begin{eqnarray}
 \nonumber {\rm Simple~Ratio,} &\R= &\ds\frac{\lam_S}{\lam_H}, \\
 \nonumber {\rm Color,} &\C= &\ds\log_{10}\left(\frac{\lam_S}{\lam_H}\right) \,, \\
 {\rm Fractional~Difference,} 
	&\HR=&\ds\frac{\lam_H-\lam_S}{\lam_H+\lam_S}.
\end{eqnarray}
These quantities are characteristics of the spectrum in question, rather 
than of the observed data, like the quantities in Equation~\ref{e:bgRCHR}.
As such, the rewritten hardness ratio is of more direct scientific interest.
Note also that while these quantities are defined entirely in terms of
expected source intensities in the different passbands, the background is
implicitly taken into account via direct modeling, as in
Equations~\ref{park:eq:SH}~and~\ref{park:eq:B}.

\subsection{Bayesian Approach}
\label{park:sec:Bayes}

As a new method of evaluating hardness ratios, we adopt a Bayesian
approach.
Bayesian analysis is explained in detail in numerous articles in the
astronomical literature (e.g., Gregory \& Loredo 1992, Kashyap \& Drake
1998, van Dyk et al.\ 2001, Kashyap et al.\ 2002, Protassov et al.\ 2002),
and here we describe it only in rudimentary terms.
In this approach, the prior knowledge for a parameter (the {\sl prior
probability distribution}, e.g., $p(\lam_S,\xi_S)$) is combined with
the information in the observed data (the {\sl likelihood}, e.g.,
$p(S,B_S|\lam_S,\xi_S)$) to derive the probability distribution
of the model parameters (the {\sl posterior distribution}, e.g.,
$p(\lam_S,\xi_S|S,B_S)$) via Bayes' theorem,\footnote{
The
notation $p(A|B)$ is to be read as ``the probability that $A$
is true {\sl given} that $B$ is true.''  Thus, except for prior
distributions, all the probability distributions that we use are
conditional distributions.}
i.e.,
\begin{eqnarray}
  p(\lam_S,\xi_S|S,B_S) = \frac{p(\lam_S,\xi_S)p(S,B_S|\lam_S,\xi_S)}
	{\int\!\!\int\!p(\lam_S,\xi_S)p(S,B_S|\lam_S,\xi_S)d\xi_Sd\lam_S} \,,
  \label{park:eq:Bayes}
\end{eqnarray}
and similarly for $p(\lam_H,\xi_H|H,B_H)$.
Because the posterior distribution in Equation~\ref{park:eq:Bayes} needs only
be computed
up to a normalizing constant, we often deal with an unnormalized
posterior density, i.e., $p(\lam_S,\xi_S|S,B_S)\propto
p(\lam_S,\xi_S)p(S,B_S|\lam_S,\xi_S)$. Bayesian statistical inferences
for a parameter are based on the posterior distribution, so that we
consider its various summary statistics; see
\S\ref{park:sec:stats} for details.

Because the underlying likelihood functions
are Poisson distributions, we assign so-called
conjugate $\gamma$-prior distributions for both source and background
intensities (van Dyk et al.\ 2001); a conjugate prior distribution
ensures that a posterior distribution follows the same parametric
form as the prior distribution, e.g., a $\gamma$-distribution is a
conjugate prior for a
Poisson likelihood.\footnote{
Formally, these are {\it semi-conjugate}
prior distributions in that they are conjugate to a particular conditional
distribution. For example, the $\gamma$-prior distribution on $\lam_S$
is conjugate to the Poisson distribution of the source counts among the 
observed soft counts, denoted $\eta_S$ in Appendix~\ref{park:eq:eta}.
The  $\gamma(\alpha,\beta)$ distribution is a continuous distribution
on the positive real line with probability density function
$$p(X)= \frac{1}{\Gamma(\alpha)} \beta^\alpha X^{\alpha-1} e^{-\beta\,X}\,,$$
and has a mean of $\alpha/\beta$, and a variance of $\alpha/\beta^2$ for
$\alpha,\beta>0$. 
}
That is, we assign
independent $\gamma$-prior distributions for the source and background
counts $\lam$ and $\xi$,
\begin{eqnarray}
  \lam_S \sim \gamma\big(\psi_{S_1},\psi_{S_2}\big) &\mbox{and}& 
  \lam_H \sim \gamma\big(\psi_{H_1},\psi_{H_2}\big) \,,  \label{park:eq:lam} \\
  \xi_S \sim \gamma\big(\psi_{S_3},\psi_{S_4}\big) &\mbox{and}& 
  \xi_H \sim \gamma\big(\psi_{H_3},\psi_{H_4}\big) \,,  \label{park:eq:xi}
\end{eqnarray}
where the values of $\psi$ are calibrated according to our prior
knowledge, or lack thereof, about the parameters; see
\S\ref{park:sec:priors} for our discussion on choosing a prior
distribution.

We describe the application of Bayes' Theorem and the derivation of
the posterior probability distribution functions for the interesting
parameters in Appendix~\ref{sec:appA}.  Here, we note that since
$\lam_S$ and $\lam_H$ are independent of each other, their joint
posterior probability distribution is the product of the individual
marginal posterior distributions,\footnote{
The marginal posterior
distribution of interest is computed by integrating out the so-called
nuisance parameters out of their joint posterior distributions. 
For example, $p(\lam_S|S,B_S)=\int p(\lam_S,\xi_S|S,B_S)\, d\,\xi_S$.}
written as (see Equation~\ref{park:eq:joint}):
\begin{equation}
  p(\lam_S,\lam_H|S,H,B_S,B_H)=p(\lam_S|S,B_S)p(\lam_H|H,B_H) \,.
\end{equation}
Then the posterior distribution of each type of hardness ratio can be
computed by transforming $\lam_S$ and $\lam_H$ to the appropriate variables
and marginalizing over the resulting nuisance variable, as follows:
\begin{equation}
p(\R|S,H,B_S,B_H)\,d\R =
d\R \int_{\lam_H} d{\lam_H}~\lam_H~p(\R\lam_H,\lam_H|S,H,B_S,B_H) \,,
\label{e:postR}
\end{equation}
\begin{equation}
p(\C|S,H,B_S,B_H)\,d\C =
d\C \int_{\lam_H} d{\lam_H}~\lam_H~10^{\C}\ln(10)~p(10^{\C}\lam_H,\lam_H|S,H,B_S,B_H)
	\,, 
\label{e:postC}
\end{equation}
\begin{equation}
p(\HR|S,H,B_S,B_H)\,d\HR =
d\HR \int_{\omega} d{\omega}~\frac{\omega}{2}~
p\bigg(\frac{(1-\HR)\omega}{2},\frac{(1+\HR)\omega}{2}\Big|S,H,B_S,B_H\bigg) \,,
\label{e:postHR}
\end{equation}
with $\omega = \lam_S+\lam_H$.  The integrals can be computed using
either Monte Carlo integration or Gaussian quadrature (throughout this
paper, we use the term ``quadrature'' to refer exclusively to the latter).
We use both methods of
integration because neither has the clear advantage over the other
in our case; see \S\ref{park:sec:limit} and Appendix~\ref{sec:appA}.

\subsubsection{Finding Point Estimates and Error Bars}
\label{park:sec:stats}

Bayesian statistical inferences are made in terms of probability
statements, and thus we consider various summary statistics for a
posterior distribution.  Here we briefly describe the standard summaries 
of location and methods of computing error bars; a fuller description 
along with detailed examples can be found in Park et al.\ (2006,
in preparation).  Commonly used summaries of location are the
mean, median, and mode(s) of the distribution: The posterior mean is
the posterior expectation of the parameter, the posterior median is
the point that divides the posterior probability evenly such that 50\%
of the probability resides below and above its value, and the
posterior mode is the most likely value of the parameter given the
data.

When a posterior simulation of size $N$ is obtained with a Monte
Carlo sampler (see \S\ref{park:sec:gibbs}), the posterior mean is the simple average of the sample
draws and the posterior median is the $[N/2]^{\rm th}$ draw after
sorting the draws in increasing order. To compute the posterior mode,
we repeatedly bisect the Monte Carlo draws and
choose the half with more draws until the range of the chosen half
is sufficiently narrow. The midpoint of the resulting narrow bin
approximates the posterior mode.

With quadrature (see
\S\ref{park:sec:quad}), we can
obtain equally spaced abscissas and the corresponding posterior
probabilities. Thus, the posterior mean is computed as the sum of the
product of an abscissa with their probabilities (i.e., the dot product
of the vector of abscissa with the corresponding probability vector).
The posterior median is computed by summing the probabilities
corresponding to the ordered abscissa one-by-one until a cumulative
probability of 50\% is reached. The posterior mode is simply the point
among the abscissa with the largest probability.

Unlike with point estimates above, there is no unique or preferred way
to summarize
the variation of a parameter.  Any interval that encompasses a suitable
fraction $\al$ of the area under the probability distribution qualifies
as an estimate of the variation.  Of these, two provide useful measures
of the uncertainty: the equal-tail posterior interval, which is the
central interval that corresponds to the range of values above and
below which lies a fraction of exactly $(\al/2)$ of the posterior
probability, is a good measure of the width of the posterior distribution;
and the highest posterior density (HPD) interval, which is the range of
values that contain a fraction $(1-\al)$ of the posterior probability,
and within which the probability density is never lower than that outside
the interval.  The HPD-interval always contains the mode, and thus serves
as an error bar on it.

For a symmetric, unimodal posterior distribution, these two posterior
intervals are identical. The equal-tail interval is invariant to
one-to-one transformations and is usually easier to compute.
However, the HPD-interval always guarantees the interval with the
smallest length among the intervals with the same posterior probability. 
Neither of these intervals is necessarily symmetric around the point
estimate, i.e., the upper and lower bounds may be of different
lengths. For consistency, we refer to such intervals as {\it posterior
intervals}; others also refer to them as confidence
intervals\footnote{
Technically, confidence intervals are defined in
terms of how frequently they contain the ``true'' value of the
estimand, rather than in terms of a posterior probability
distribution.} or credible intervals.

For a Monte Carlo simulation of
size $N$, we compute either the equal-tail interval or an interval
that approximates the HPD interval. (Unless otherwise stated,
here we always quote the equal-tail interval for the Monte Carlo
method and the HPD-interval for the quadrature.)
The equal-tail posterior interval in this case is computed by choosing
the $[(\al/2)N]^{\rm th}$ and the $[(1-\al/2)N]^{\rm th}$ draws as
the boundaries.  An approximate HPD-interval is derived by comparing
all intervals that consist of the $[X]^{\rm th}$ and the
$[X+(1-\al)N]^{\rm th}$ draws and choosing that $X$ which gives the
shortest length among them.

When the posterior density is computed by the
quadrature, we split parameter space into a number of bins and
evaluate the posterior probability at the midpoint of each bin. In
this case, a $100(1-\al)\%$ HPD-interval can be computed by beginning
with the bin with the largest posterior probability and adding
additional bins down to a given value of the probability density
until the resulting region contains at least a fraction
$(1-\al)$ of the posterior probability.

\section{Verification: Simulation Study}
\label{park:sec:verify}

In order to compare the classical method with our Bayesian method, we
carried out a simulation study to calculate coverage rates of the
classical and posterior intervals.  Given pre-specified values of the
parameters, source and background counts were generated and then used
to construct 95\% classical and posterior intervals of each hardness
ratio using the methods discussed in this article.  From the simulated
data we calculated the proportion of the computed intervals that
contain the true value of the corresponding hardness ratio. (This is
the coverage rate of the classical and probability intervals.)  In the
ideal case, 95\% of the simulated intervals would contain the true
value of each hardness ratio.  Besides the coverage rate, the average
length of the intervals and the mean square error of point estimates
were also computed and compared. The mean square error of the point
estimate $\hat{\th}$ of $\th$ is defined as the sum of the variance
and squared bias for an estimator, i.e., ${\rm
MSE}(\hat{\th})=\E[(\hat{\th}-\th)^2]=\Var(\hat{\th})+[\E(\hat{\th})-\th]^2$
A method that constructs shorter intervals with the same coverage rate
and produces a point estimate with a lower mean square error is
generally preferred. The entire simulation was repeated with different
magnitudes of the source intensities, $\lam_S$ and
$\lam_H$. Intrinsically, we are interested in the following two
prototypical cases:
\begin{mathletters}
\begin{description}
  \item[\kase~I :] hardness ratios for high counts, i.e.,
    \begin{eqnarray} 
	\lam_S=\lam_H=30,\ \xi_S=\xi_H=0.1;
	\label{e:case1}
    \end{eqnarray}
  \item[\kase~II :] hardness ratios for low counts, i.e., 
    \begin{eqnarray} 
	\lam_S= \lam_H=3,\ \xi_S=\xi_H=0.1.
	\label{e:case2}
    \end{eqnarray}
\end{description}
\end{mathletters}
In both cases we adopt a background-area--to--source-area ratio
of $r=100$ (see Equation~\ref{park:eq:B}), i.e., we take the
observed counts in the background region to be 10.
Note that these are written with reference to the counts observed
in the ``source region'', i.e., the units of $\lam_S, \lam_H, \xi_S,
\xi_H$ are all [ct~(source area)$^{-1}$], and that we have set
$\eff_S=\eff_H=1$ here.  The actual extent of the
source area is irrelevant to this calculation.

\begin{figure}[t]
\centerline{\includegraphics[width=4.5in,angle=0]{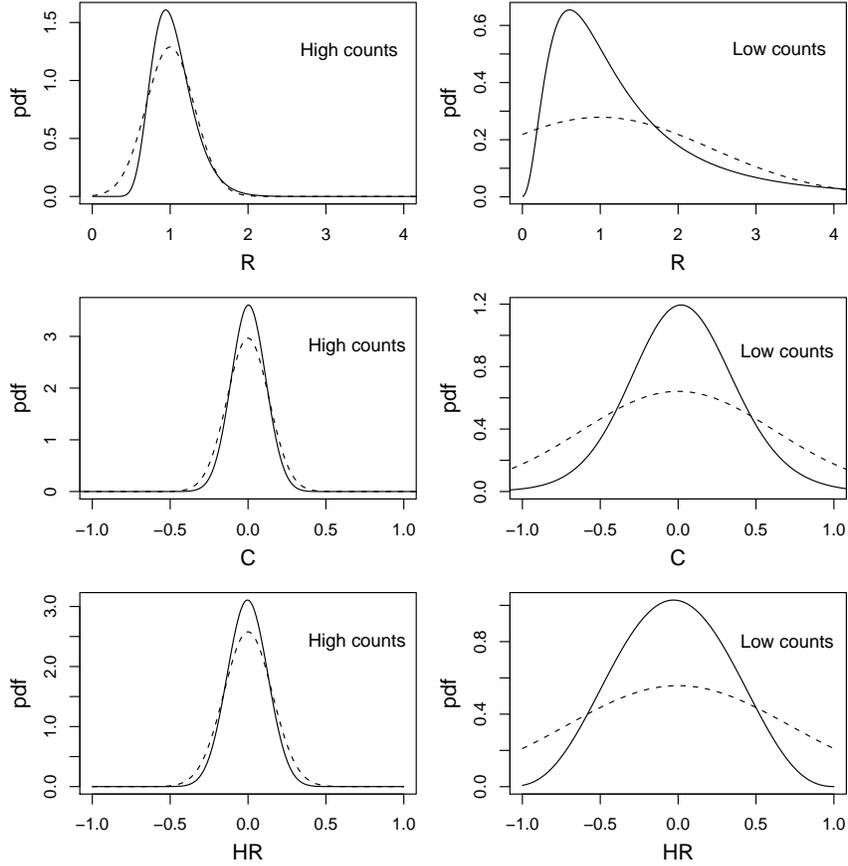}}
\caption{Comparison of the shape of posterior distribution for
	hardness ratios $\R$, $\C$, and $\HR$ with the classical
	Gaussian approximation for high counts 
        (\kase~I; left column of the figure) and 
        for low counts (\kase~II; right column of the figure).
	The solid lines represent the posterior distributions of 
	the hardness ratios and the dashed lines a Gaussian distribution
	with mean and standard deviation equal to the classical estimates.
	A prior distribution that is flat on the real line
	($\phi=1$, see \S\ref{park:sec:priors}) is adopted for
	the Bayesian calculations.  Note that as the number of counts
	increase, the two distributions approach each other.  At
	low counts the two distributions differ radically, with
	the Classical distributions exhibiting generally undesirable
	properties (broad, and extending into unphysical regimes).
	\label{park:fig:comp}}
\end{figure}

This simulation study illustrates two typical cases, i.e., 
high counts and low counts sources: \kase~I represents high 
counts sources for which Poisson assumptions tend to
agree with Gaussian assumptions; \kase~II represents low counts 
sources where the Gaussian assumptions are inappropriate.
In \kase~I, the posterior distributions of the hardness ratios 
agree with the corresponding Gaussian approximation of the
classical method, but in \kase~II, the Gaussian assumptions
made in the classical method fail.  This is illustrated in
Figure~\ref{park:fig:comp}, where we compare the two methods
for a specific and particular simulation:
we assume $S=H=30$ in \kase~I and $S=H=3$ in \kase~II, compute
the resulting posterior distributions of hardness ratios using
the Bayesian method, and compare it to a Gaussian distribution with mean and
standard deviation equal to the classical estimates.
The right panels in Figure~\ref{park:fig:comp} show that
there is a clear discrepancy between the two methods.

\clearpage
\clearpage

\begin{table}[t]
  \begin{center}
    \caption{Statistical Properties of Bayesian and Classical Methods.}
  \medskip
  \begin{tabular}{ccccccc} \hline\hline
  & & Hardness & Coverage & Mean & \multi{Mean Square Error}\\
  \cline{6-7}
  & {\up{Method}} & Ratio & Rate & Length & of mode & of mean\\
  \hline
  \divi{}&                 & $\R$  & 95.0\% & 1.24 &\multi{0.065}\\
  \divi{}&{\up{Classical}} & $\C$  & 99.0\% & 0.54 &\multi{0.013}\\
  \divi{}&{\up{method}}    & $\HR$ & 98.5\% & 0.61 &\multi{0.016}\\
  \cline{2-7}
  \divi{}&                 & $\R$  & 96.0\% & 1.07 & {\bf 0.056} & 0.069\\
  \divi{\up{\kase~I}}&{\up{Monte Carlo}} & $\C$  & 96.0\% & 0.44 & {\bf 0.012} & 0.012\\
  \divi{\up{(high counts)}}&{\up{method}}   & $\HR$ & 96.0\% & 0.49 & 0.016 & {\bf 0.015}\\
  \cline{2-7}
  \divi{}&                 & $\R$  & 94.5\% & 1.03 & {\bf 0.057} & 0.069\\
  \divi{}&{Quadrature}     & $\C$  & 96.0\% & 0.43 & {\bf 0.012} & 0.012\\
  \divi{}&{}               & $\HR$ & 94.5\% & 0.49 & 0.016 & {\bf 0.015}\\
  \hline\hline
  \divi{}&                 & $\R$  & 97.5\%  & 192.44 &\multi{93.27}\\
  \divi{}&{\up{Classical}} & $\C$  & 100.0\% & 6.02   &\multi{0.27} \\
  \divi{}&{\up{method}}    & $\HR$ & 100.0\%  & 3.70   &\multi{0.21} \\
  \cline{2-7}
  \divi{}&                 & $\R$  & 98.0\% & 15.77 & {\bf 0.328} & 85.482\\
  \divi{\up{\kase~II}}&{\up{Monte Carlo}} & $\C$  & 98.0\% & 1.54  & {\bf 0.078} & 0.113  \\
  \divi{\up{(low counts)}}&{\up{method}}   & $\HR$ & 98.0\% & 1.26  & 0.181 & {\bf 0.083}  \\
  \cline{2-7}
  \divi{}&                 & $\R$  & 97.0\% & 8.18 & {\bf 0.394} & 20.338\\
  \divi{}&{Quadrature}     & $\C$  & 99.5\% & 1.51 & {\bf 0.074} & 0.112 \\
  \divi{}&{}               & $\HR$ & 95.0\% & 1.23 & 0.187 & {\bf 0.083} \\
  \hline\hline
  \end{tabular}
  \label{park:tbl:coverage}
  \end{center}
\end{table}

\begin{figure}[htb!]
\centerline{\includegraphics[width=6.5in,angle=270]{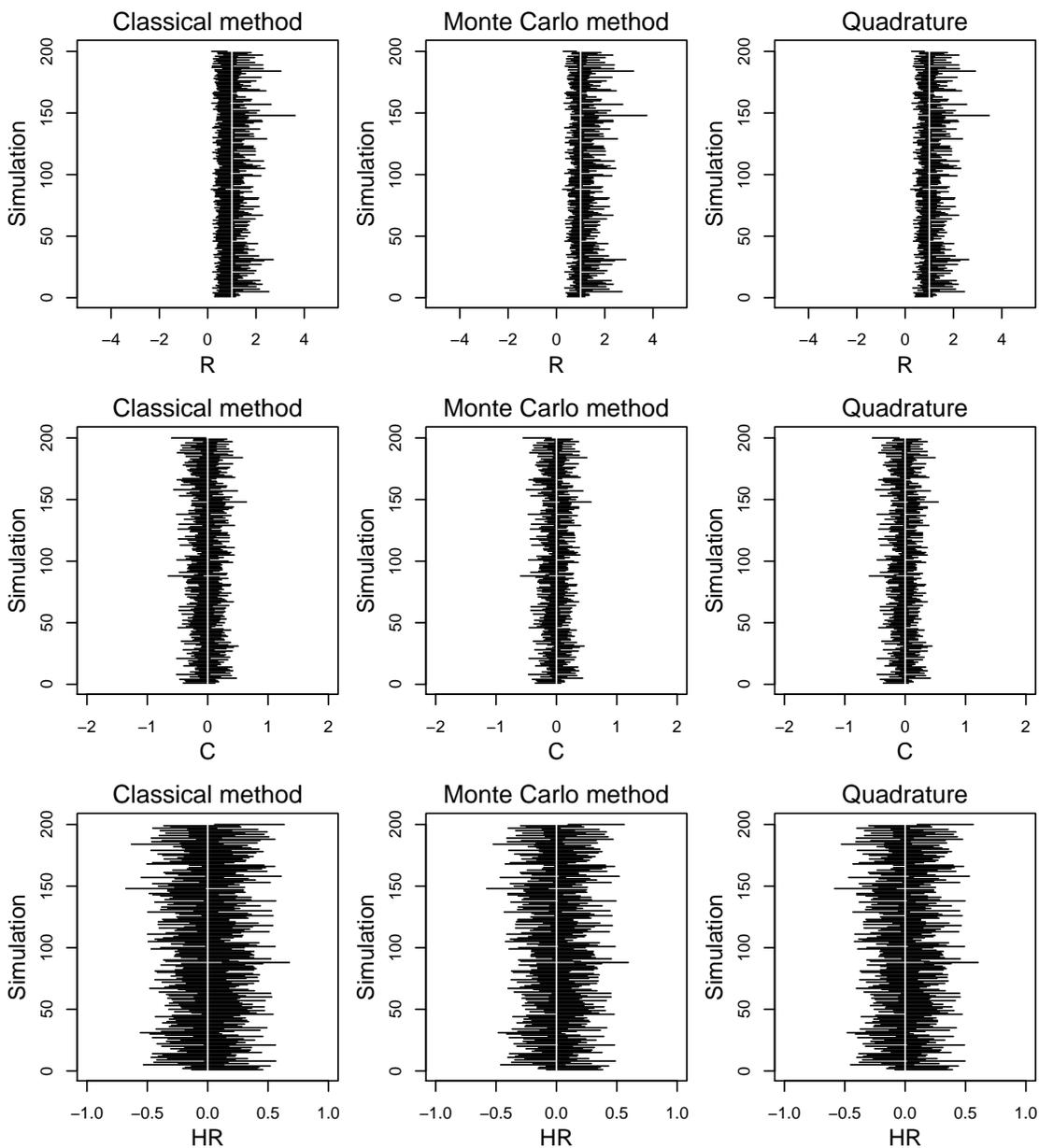}}
\caption{Simulated range of hardness ratios calculated using three
different methods in the high counts case (\kase~I;
see Equation~\ref{e:case1}) -- the classical method ({\sl left}),
Monte Carlo method ({\sl middle}), and quadrature ({\sl right}) --
and for the three types of hardness ratios -- $\R$ ({\sl top}),
$\C$ ({\sl middle}), and $\HR$ ({\sl bottom}).  The horizontal
lines are the 95\% intervals computed for each simulated set
of counts, and the vertical white line in each panel represents
the true value of the hardness ratio.  Note that all the different
methods exhibit similar performance in this case.
\label{park:fig:case1} }
\end{figure}

\begin{figure}[htb!]
\centerline{\includegraphics[width=6.5in,angle=270]{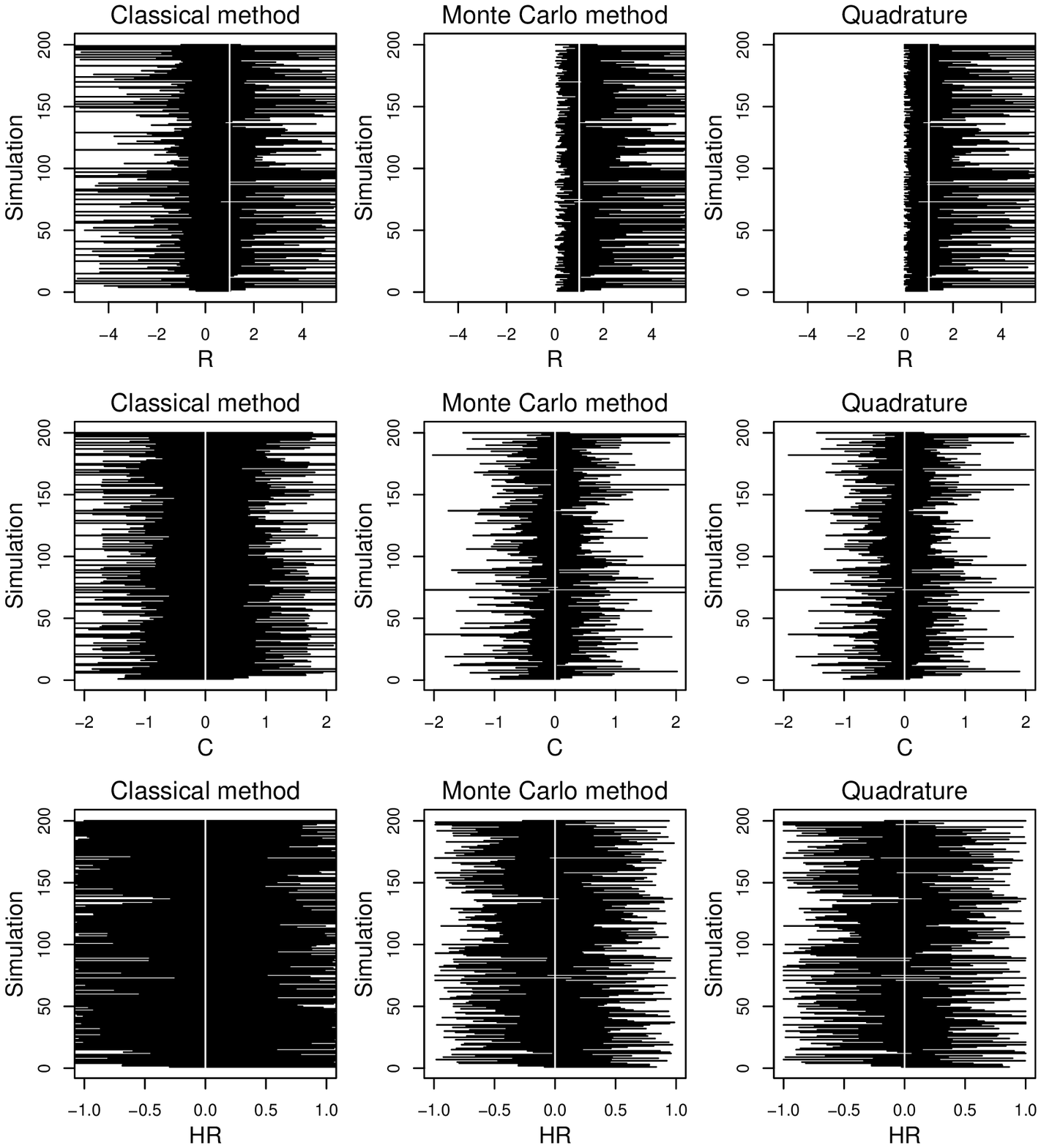}}
\caption{As in Figure~\ref{park:fig:case1}, for simulated
range of hardness ratios in the low counts case (\kase~II;
Equation~\ref{e:case2}).  In this, the low counts case, the
the Bayesian methods dramatically outperform the classical method.
\label{park:fig:case2} }
\end{figure}

In order to compare the statistical properties of estimates based on
the Bayesian and classical methods, we simulate test data under
both \kase~I and \kase~II.  Table~\ref{park:tbl:coverage}
presents the results of 200 simulated sources for each of the two
cases. The data are used to compute point estimates and 95\% error
bars by using the classical method, Monte Carlo sampler,
and numerical integration by quadrature. The Bayesian methods use
non-informative prior distributions on $\lam$ and $\xi$; in particular,
$\lam\sim\gamma(1,0)$ and $\xi\sim\gamma(0.5,0)$, see \S\ref{park:sec:priors}.
The results of \kase~I indicate that all
three methods are comparable although the quadrature
yields slightly shorter intervals and point estimates with a smaller
mean square error. Figure~\ref{park:fig:case1} illustrates that the
resulting intervals from the methods are almost identical. On the
other hand, the results of \kase~II appear in
Figure~\ref{park:fig:case2} and confirm that the Bayesian methods
outperform the classical method because the mean length of the
classical intervals are wider than the posterior intervals, and the
classical estimates exhibit much larger mean square error.  For
example, the hardness ratio $\HR$ by definition is limited to the
range $[-1,+1]$, so that the maximum length of the intervals should be
two. However, the mean length of simulated classical intervals is
3.7, which is clearly larger than needed. Since
quadrature tends to yield estimates with smaller mean square errors
and shorter posterior intervals while maintaining high coverage rates,
it is generally preferred to the Monte Carlo method. However, this comes at
a price: because the summation inside the posterior density is over
the detected source counts, quadrature is
computationally more expensive as the source counts increase; on the
other hand, the Monte Carlo method is much faster regardless of the number
of source counts.  Thus, we recommend using quadrature with
relatively low counts data (e.g., $S<20$ or $H<20$) and the Monte Carlo method
with higher counts.

Because the Bayesian inference for each hardness ratio is based on
its posterior distribution, we consider these two point estimates, the
posterior mode and the posterior mean. Table~\ref{park:tbl:coverage}
presents a comparison of the two proposed estimates in terms of their
mean square errors. In the case of $\HR$, the posterior mean seems to
be more robust to the shape of a posterior distribution than a
posterior mode because both left and right limits of $\HR$ are
bounded. For $\R$ and $\C$, on the other hand, the posterior mode
appears to perform better. Thus, the posterior mode
is preferable for both $\R$ and $\C$ and the posterior mean for $\HR$, as
shown in Table~\ref{park:tbl:coverage}.

\section{Applications}
\label{park:sec:apply}

\subsection{Quiescent Low-Mass X-ray Binaries}
\label{park:sec:qLMXB}

Low-mass X-ray binaries (LMXBs) are binary systems composed of a
neutron star or black-hole and a low mass ($<1-2~M_{\odot}$)
donor. These binaries are generally transient sources and spend most of their
time in quiescence (qLMXB's).  However, even when not active they emit X-rays
due to thermal emission from the surface of the neutron star, and
sometimes due to an additional hard component associated with residual
low-level accretion (e.g., Campana et al.\ 1998, Brown, Bildsten, \&
Rutledge 1998).  Identification and study of these
quiescent systems is very important since they provide a direct census
of the overall LMXB population (and therefore the population of neutron
stars), and also constrain their duty cycles.
Although only the identification of their optical counterparts can
unambiguously classify an X-ray source as a qLMXB, soft X-ray
spectra or hardness ratios consistent with those expected for thermal
emission by neutron stars are very useful for identifying candidate
qLMXB's (Heinke et al.\ 2003).

\begin{figure}[p]
\centerline{\includegraphics[width=4in,angle=270]{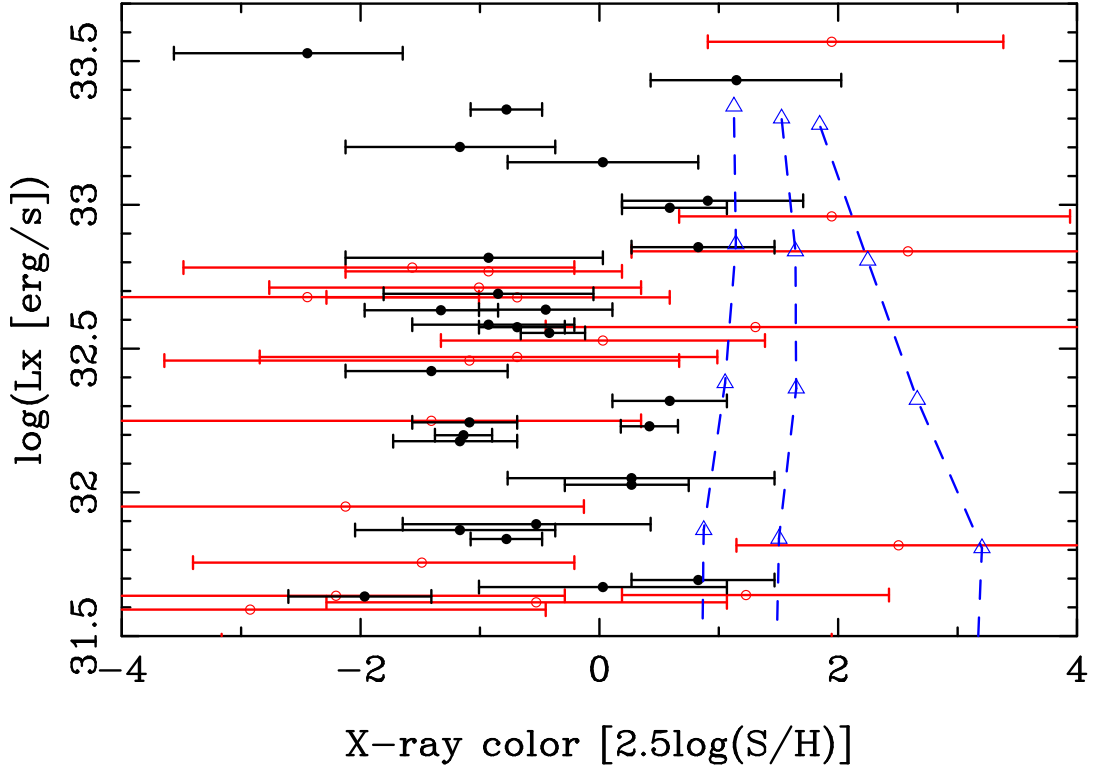}}
\caption{The X-ray "color-magnitude" diagram for globular
cluster Terzan-5.  For each of the detected sources, a form of color
$\C=\log_{10}(\lambda_{S}/\lambda_{H})$, scaled by a factor 2.5 for
similarity with optical colors, is plotted along the abscissa, and
the $\log_{10}$(Luminosity [ergs~s$^{-1}$]) in the 0.5-6~keV band,
along the ordinate. The modes of the posterior probability functions
(computed using quadrature) for all sources are shown, along with
the 95\% confidence intervals (the error bars on the luminosity are
not plotted). The sources which would result in unreliable estimates
if the color were computed with the classical method (background
subtracted source counts $<5$ in at least one band) are marked
with open circles (red), and the remainder are marked with filled
circles. The dashed (blue) lines represent the color-luminosity
tracks expected for neutron stars with black-body spectra of
different temperatures (the open triangles correspond to temperatures
of $\log_{10}(T [{\rm K}])$ = 6.3, 6.2, 6.1 and 6.0 from top to bottom),
and varying contribution from a power-law component (slope
$\Gamma=1.5$; pure black-body for the rightmost curve, 10\% and
20\% contribution from a power-law component in the 0.5-6~keV band
luminosity for the middle and left curves respectively). For more
details on these spectral models see Heinke et al.\ (2006). 
\label{f:terzan5} }
\end{figure}

We apply the methods described in \S\ref{park:sec:model} on
a long (40~ks) \chandra\ ACIS-S observation of the globular
cluster Terzan-5 (ObsID 3798; PI: R.Wijnands).
A total of 49
sources were detected within the half-mass radius of the cluster
(for details on the observation, reduction, and analysis, see
Heinke et al.\ 2006, in preparation).
We performed the source photometry in the 0.5-2.0~keV ({\sl soft})
and 2.0-6.0~keV ({\sl hard}) bands.  In Figure~\ref{f:terzan5} we present a
color-luminosity diagram of all the detected sources
(note that the color here is scaled by a factor 2.5 to be similar
to the definition of an optical colors such as $B-V$).
The posterior modes of the color for all detected sources, computed
here using the Jeffrey's prior ($\phi=1/2$), along with the 95\%
highest posterior density intervals, are shown.  Those sources which
are not amenable to analysis via the classical method (weak sources
with background-subtracted source counts in at least one band being
$<5$) are marked with open circles, while the remainder are
marked with filled circles.
Thermally emitting neutron stars are expected to have X-ray colors
in the range $[0.5,2.5]$ (indicated by dashed lines in the figure).
The softer limit (dashed curve to the right; color~$\approx2$)
corresponds to a purely thermal emission spectrum, and the harder
limit (dashed curve to the left; color~$\approx0.5$) corresponds
to a 20\% contribution from a
power-law of index $1.5$.
From this figure it is clear that the Bayesian method, apart from
providing more accurate confidence intervals for the source colors,
allows us to estimate hardness ratios for the sources which were
undetected in one of the two bands.
We find 15 sources whose error bars overlap this region and are
thus candidate qLMXB's.  Note that 8 of these candidates could not
have been so classified using the classical method because there
are too few source counts.
Conversely, many sources that lie at the edge of the region of interest
would have been mistakenly classified as candidate qLMXB's with the
classical method if the overly large error bars it produces had been
used (see discussion of coverage rates in Table~\ref{park:tbl:coverage}).

\clearpage

\subsection{Evolution of a Flare}
\label{park:sec:flare}

In the previous section (\S\ref{park:sec:qLMXB}), we demonstrated
the usefulness of using
the full Bayesian machinery in classifying weak sources in comparison
to the classical method.  Here, we shall go further and show that
new avenues of analysis and interpretation are opened up when we
take advantage of the full probabilistic description available to us.
Such analysis methods are difficult to implement correctly and
generally cannot be justified under the classical method.  As a
specific illustrative example, we consider the time evolution of
a large X-ray flare on a low-mass dwarf, Ross\,154 (Wargelin et
al.\ 2006).

Stellar coronae consist of magnetically confined hot plasma
($T\sim{1-10}$~MK) in the outer atmospheres of late-type stars,
and emit optically thin thermal emission composed of Bremsstrahlung
continuum and atomic line emission components.  Detailed analysis
of these line-rich spectra yield valuable clues regarding the
temperature and emission structure on the star, its composition,
and the processes that drive coronal heating (see e.g., Rosner,
Golub, \& Vaiana, 1985).  Of special importance are flares, which
result from impulsive energy input into the corona by the reconnection
of highly strained magnetic flux tubes, and provide additional
information on the dynamics of the corona.  Analysis of flare
spectra during decay yields constraints on the physical site and
size of the flare loops and on the heating process (see, e.g.,
van den Oord \& Mewe 1989, Pallavicini, Tagliaferri, \& Stella 1990,
Serio et al.\ 1991, Schmitt \& Favata 1999).

A useful technique developed explicitly to analyze hydrodynamically
evolving loops is to track their evolution in density-temperature
space (Reale, Serio, \& Peres 1993, Reale et al.\ 1997, Reale
et al.\ 2004).  This makes possible the modeling of flares to derive
physically meaningful parameters such as the length of the loop, the
heating function, etc.
Unfortunately, a comprehensive analysis along these lines requires
that complex model spectra be fit to the data at each stage of the
flare evolution, and because fitting requires a large number of
counts (generally $>500$) to produce reliable results, the resolution
at which the flare evolution can be studied is strongly limited.
This is especially so during the rising phase of flares, where the
physically most interesting processes are of short duration and
the observed counts are also small.  
However, hardness ratios can be constructed at relatively finer
time binning resolution and can be used as a proxy for the spectral
fit parameters.
In the following, we apply the Bayesian method described above
(see \S\ref{park:sec:model}) to \chandra\ data of a
stellar X-ray flare, and demonstrate its usefulness.

\begin{figure}[htb!]
\centerline{\includegraphics{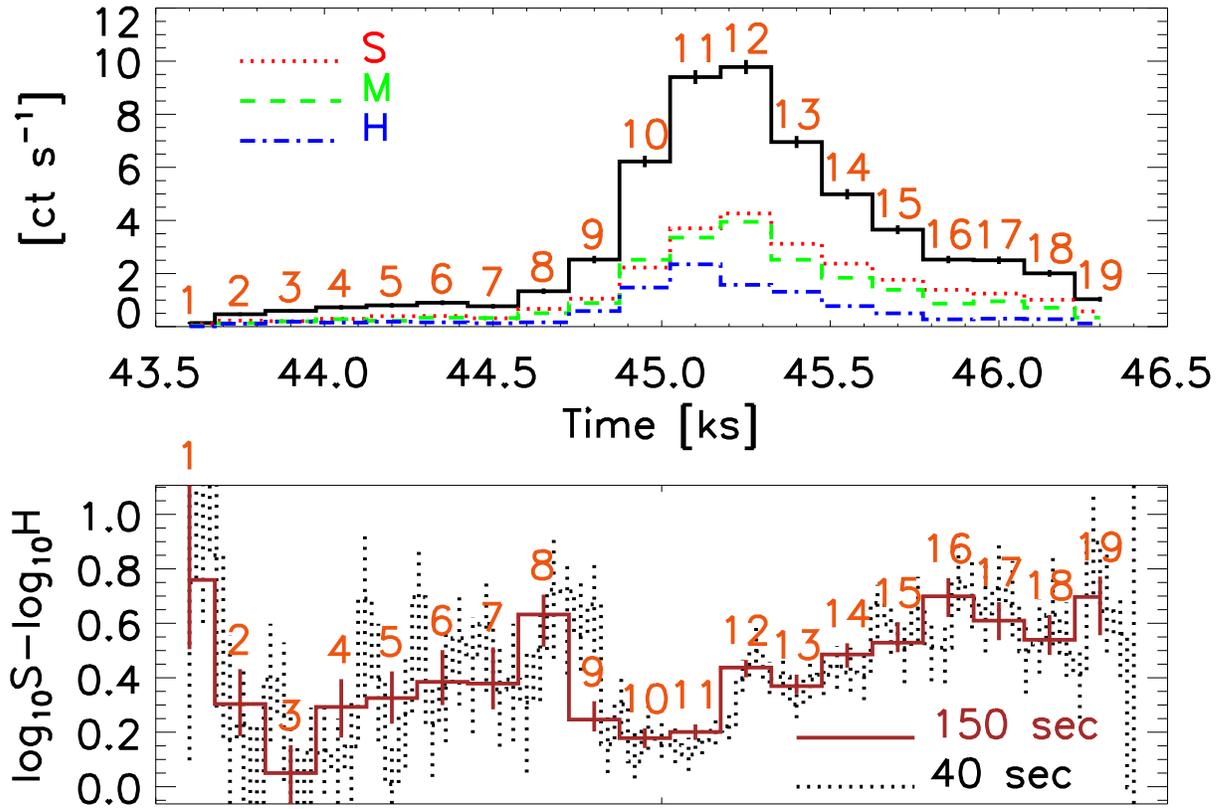}}
\caption{
{\sl Upper Panel:} Light-curves of Ross\,154.  The observed
count rate in the source is shown as a function of time since
the beginning of the observation, for counts in the {\sl soft}
(dotted histogram), {\sl medium} (dashed histogram), {\sl hard}
(dash-dotted histogram), and the total (solid histogram).  A
bin size of 150~s was chosen for convenience, and the sequence
of bins are marked along the light curve for future reference.
{\sl Lower Panel:} Hardness ratio evolution of Ross\,154.  The
color, $\C=\log_{10}\frac{S}{H}$ is plotted for a bin size of
$\delta{t}=150$~s (solid histogram) and $\delta{t}=40$~s
(dotted histogram), along
with the associated error bars.  Larger values of $\C$ indicate
a softer spectrum, and vice versa.
} \label{f:colmag_evol_lc}
\end{figure}

A large flare was observed on Ross\,154, a nearby (2.93 pc) active
M dwarf (M4.5Ve), during a 57~ks observation with the \chandra/ACIS-S
detector (ObsID 2561; PI: B.Wargelin).  During the flare, the
counting rate increased by over a
factor of 100.  Because of strong pileup in the core of the PSF,
only those data within an annular region surrounding the center
can be used. (Details of the reduction and analysis are given in
Wargelin et al.\ 2006.)  The light-curve of the
source during the flare is shown in Figure~\ref{f:colmag_evol_lc},
along with light-curves for smaller passbands ({\sl soft}, 0.25-1~keV;
{\sl medium}, 1-2~keV; {\sl hard}, 2-8~keV).  Note that the {\sl hard} band
light-curve peaks at an earlier time than the softer bands, an impression
that is borne out by the evolution of the color, $\C$ (lower panel of
Figure~\ref{f:colmag_evol_lc}).

\begin{figure}[htb!]
\centerline{\includegraphics{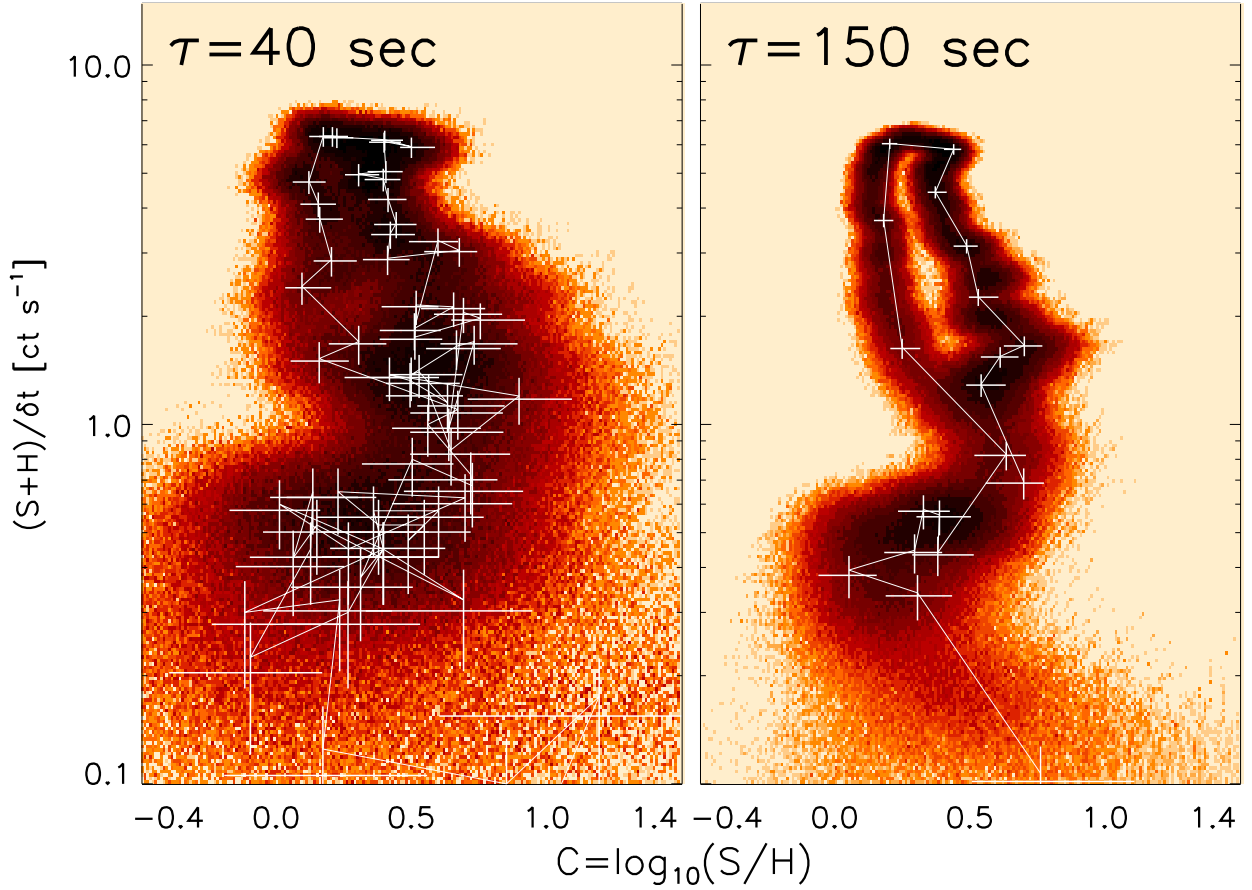}}
\caption{Color-intensity evolutionary tracks.  Spectral hardness
increases to the left and source brightness increases towards the
top.  The paired color and
intensity points (shown separately in Figure~\ref{f:colmag_evol_lc})
are plotted for bin sizes of 40~s ({\sl left}) and 150~s ({\sl right}).
The error bars for each quantity are also plotted as horizontal and
vertical bars.  The points are connected by a thin solid line to
aid in visualization of the temporal sequence (see also
Figure~\ref{f:colmag_evol_all}).  Samples of $(\lam_S,\lam_H)$
obtained from their posterior probability distributions
for each bin and for various starting time phases are
displayed as the shaded image underneath the evolutionary tracks.
	\label{f:colmag_evol_cyc}}
\end{figure}

The same data can also be represented in a color-intensity diagram,
as in Figure~\ref{f:colmag_evol_cyc}, which shows scatter plots of
the color and intensity at each time bin along with their computed
errors (connected by a thin line to indicate the time sequence).
The advantage of doing so is
that color is inversely correlated with plasma temperature, and the
intensity approximately tracks the plasma density.\footnote{
X-ray luminosity,
$L_X \propto EM \cdot P(T)$, where $EM = n_e^2 V$ is the emission
measure of coronal plasma with an electron density $n_e$ occupying
a volume $V$, and $P(T)$ is the power emitted by the plasma at
temperature $T$.  $P(T)$ is a weak function of $T$ over the
temperature range 6-20 MK expected in this source, and assuming
that the volume of emission remains constant, the observed intensity
is proportional to the square of the electron density.  For a
more detailed modeling of the $L_X-T$ conversion, see Wargelin et al.\
(2006).
}
A hysteresis-like evolution is discernible in the color-intensity
track, even at a time resolution as small as 40~s.
However, the error bars on the data
points are large enough to obscure this structure at small time bin
sizes, whereas at larger bin sizes, the number of data bins are smaller
and it becomes difficult to discern the totality of the evolutionary
track, even though the pattern is persistent as bin size and starting
times are varied.  Mainly, the choices of the bin size and the phase
(i.e., the starting point) is completely arbitrary and the choice
of no single value can be properly justified.  However, Bayesian
analysis provides a simple method to work around this difficulty,
by simply considering the bin size as a nuisance parameter of the
problem and to marginalize over it.

\begin{figure}[hbt!]
\centerline{\includegraphics{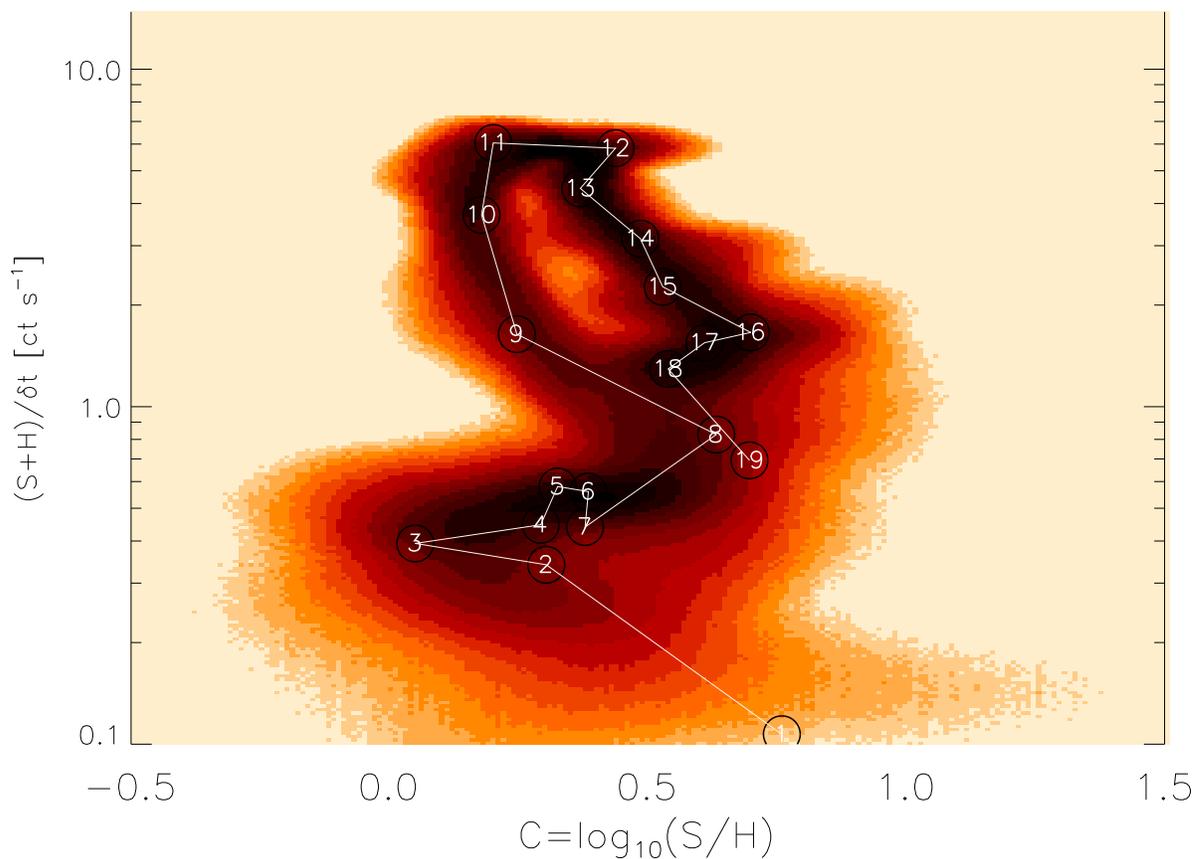}}
\caption{Ross\,154 color-intensity track.
As in Figure~\ref{f:colmag_evol_cyc}, but for an image obtained
by averaging cycle-spun images at various bin sizes ranging
from 40~s to 400~s.  The persistence of the evolutionary track
of the flare on Ross\,154 is quite clear in this visualization.
The points corresponding to each bin of the light-curve obtained
at a bin size of 150~s are superimposed on the image, with a
white solid line connecting them in temporal sequence.
A clear pattern of the flaring plasma evolution can be seen, first
rising in temperature and intensity, softening at the peak, and then
decaying in temperature and intensity.
}
	\label{f:colmag_evol_all}
\end{figure}

Thus, in order to ameliorate the effects of choosing a single time
bin size
and a starting time, we have carried out a series of calculations
using the Monte Carlo method: we obtain different light-curves by varying
the phase of the binning (i.e, the starting time), and for each
light-curve thus obtained, we obtain 2000 samples of $(\lam_S,\lam_H)$
for each data point.  We hold the time bin size fixed during this
process.  After cycling
through all the phases, the samples of $(\lam_S,\lam_H)$ are all
combined to form a single image (the shaded images in
Figure~\ref{f:colmag_evol_cyc}).  This procedure, also called
``cycle-spinning'' (Esch 2003), allows us to discern the pattern of
the evolutionary track clearly.
The longer the source spends at any part of the color-intensity
diagram, the darker the shade in the images, and the statistical
error in the calculation is reflected by the gradient of the
shading.
Further, all
the cycle-spun images thus constructed, for bin sizes ranging from
40~s to 400~s, were then averaged to produce a single coherent image,
thus effectively marginalizing over the bin sizes; such an image
includes all the errors associated with the analysis
(Figure~\ref{f:colmag_evol_all}).

Note that this type of analysis is not possible using the
classical method.  For instance, in the Bayesian case, the
samples are drawn from the posterior probability distributions
of $S$ and $H$.  No such sampling technique is available in
the classical case,\footnote{
New counts realizations can be obtained by bootstrapping
from the observed counts, but such sampling will be quantized
and will render the shaded images unusable at lower counts
intensities.
}
and naively using the Gaussian approximations as in
Equation~\ref{park:eq:sig2} will overestimate the errors
and wash out the signal (see \S\ref{park:sec:verify}).
Furthermore, the concept of marginalizing over the nuisance
parameter of time bin sizes has no analog in the classical
system, and cannot be justified; and cycle-spinning leads to
correlations between the estimates at different bins since
to the statistical independence of the data will be lost.

A sharp initial hardening of the spectrum, coincident with the onset
of a small optical flare (Wargelin et al.\ 2006) is visible at the
beginning of the time sequence (points 1-3; see
Figures~\ref{f:colmag_evol_lc},\ref{f:colmag_evol_cyc}).
A point worth noting is that the Bayesian analysis prevents
overinterpretation of the results: at time point 3, the standard
analysis indicates that the spectrum is at its hardest, suggesting
either a non-thermal origin to the emission or significant transient
absorption, possibly due to a coronal mass ejection, but the shaded
image demonstrates that the statistical errors are sufficiently
large that a more mundane low-density thermal expansion is quite
plausible.
This stage
is followed by a rapid softening (points 3-8), interpreted as the
thermalization of the non-thermal hard X-ray flare.
Then the star, which
lies in the lower central portion of the color-intensity diagram
before the flare, moves
to the upper left as the flare is set off (the spectrum hardens
as intensity increases; points 8-11), turns right at flare peak
(softens as the deposited energy cascades to lower temperatures; points
11-12), and then moves down the right flank back to its original
state (decays in temperature and intensity; points 12-19).
The physical consequences of this analysis are discussed in detail by
Wargelin et al.

\clearpage

\section{Discussion}
\label{park:sec:discuss}

\subsection{Informative and Non-informative Prior Distributions}
\label{park:sec:priors}

A major component of Bayesian analysis is the specification of
a prior distribution, the probability distribution assigned
{\sl a priori} to the parameters defining the model.  This is
often deemed controversial because of the apparent subjectivity
in the assignment: if different people assign different priors
and obtain different results, how is it possible to evaluate
them?  The answer to this conundrum is to realize that all
problems always include a bias brought to the analysis by the
researcher, either in the choice of statistic, or even in the choice
of analysis method, and the specification of a prior distribution
codifies and makes explicit this bias.  Furthermore, if indeed prior
information about the value or the range of the parameter is
available, the prior allows a simple and principled method by
which this information can be included in the analysis.  If
no prior information is available, a so-called {\sl non-informative}
prior distribution must be chosen in order to not bias the result.

In our case, if there is a strong belief as to the value or range of
the hardness ratio, we can incorporate this via an informative prior
distribution, encoded as a $\gamma$-distribution (see van Dyk et al.\
2001).  In addition, the analysis of previously acquired data will
produce a posterior probability distribution that can be used as an
informative prior on future observations of the same source.

When no prior information is available, we normally use a
non-informative prior distribution. Since a Poisson
intensity is necessarily positive, three types of non-informative
prior distributions immediately present themselves: when $X|\th\sim\Pois(\th)$,
\begin{mathletters}
\begin{enumerate}
\item a non-informative prior distribution on the original scale,
\begin{equation}
	p(\th)\propto1 \,,
\label{e:prior_standard}
\end{equation}
\item a Jeffrey's non-informative prior distribution,
\begin{equation}
	p(\th)\propto I_\th^{1/2} \,, ~{\rm and}
\label{e:prior_jeffrey}
\end{equation}
\item a non-informative prior distribution under a log transformation,
\begin{equation}
	p(\log\th)\propto1 ~{\rm or~equivalently}~p(\th)\propto \frac{1}{\th} \,,
\label{e:prior_log}
\end{equation}
\end{enumerate}
\end{mathletters}
where $I_\th=\E[-\partial^2\log p(X|\th)/\partial\th^2|\th]$ is the
expected Fisher information (Casella \& Berger 2002).
The first non-informative prior distribution (Equation~\ref{e:prior_standard})
is flat from 0 to $\infty$; the second 
(Equation~\ref{e:prior_jeffrey}) is proportional to the square root
of the Fisher information; and the third 
(Equation~\ref{e:prior_log})
is flat on the whole real line
under a log transformation. The functional forms of these prior
distributions can be generalized into $p(\th)\propto\th^{\phi-1}$, i.e.,
$\th\sim\gamma(\phi,0)$ where $\phi$ is an index that varies from $0$
to $1$: the first corresponds to $\phi=1$, the second to
$\phi=1/2$; and the third to $\phi=0^+$ (where this notation
indicates that $\phi>0$, but arbitrarily close to zero, for reasons described
below).  We note that these
non-informative prior distributions are {\sl improper}, i.e.,
are not integrable.  If an improper prior distribution causes a
posterior distribution to be improper as well, then no inferences
can be made using such non-integrable distributions.  In our case,
as long as $\phi$ is strictly positive, a posterior distribution
remains {\sl proper}, i.e., integrable.  Hence, in the third case,
we adopt values of $\phi$ that are strictly positive but close in
value to $0$, e.g., $\phi=0.01$ or $\phi=0.1$.

Note that while these prior distributions are non-informative, in the
sense that in most cases they do not affect the calculated values of
the hardness ratios (see Appendix~\ref{sec:appB}), they do codify some specific
assumptions regarding the range of values that it is believed
possible.  For instance, if a large dynamic range is expected in the
observations, a flat prior distribution in the log scale is more
appropriate than a flat prior distribution in the original scale.  The
choice of the prior distribution is dictated by the analysis and must
be reported along with the results.

\subsection{$\R$ versus $\C$ versus $\HR$}
\label{park:sec:HR}

At low counts, the posterior distribution of the counts ratio,
$\R$, tends to be skewed because of the Poissonian nature of data;
$\R$ only takes positive values. The color, $\C=\log_{10}\R$,
is a log transformation of $\R$, which makes the skewed distribution more
symmetric. The fractional difference hardness ratio,
$\HR=(1-\R)/(1+\R)$, is a monotonically decreasing transformation of $\R$, such
that $\HR\rightarrow+1$ as $\R\rightarrow0$ (i.e., a source gets harder)
and $\HR\rightarrow-1$ as $\R\rightarrow\infty$ (i.e., a source gets softer).
The monotonic transformation results in a bounded range of $[-1,+1]$ for $\HR$,
thereby reducing the asymmetry of the skewed
posterior distribution.
$\R$ and $\HR$ are bounded on one side or two sides, while
$\C$ is unbounded due to the log transformation.

\begin{figure}[htb!]
\centerline{
\includegraphics[width=5.8in,angle=270]{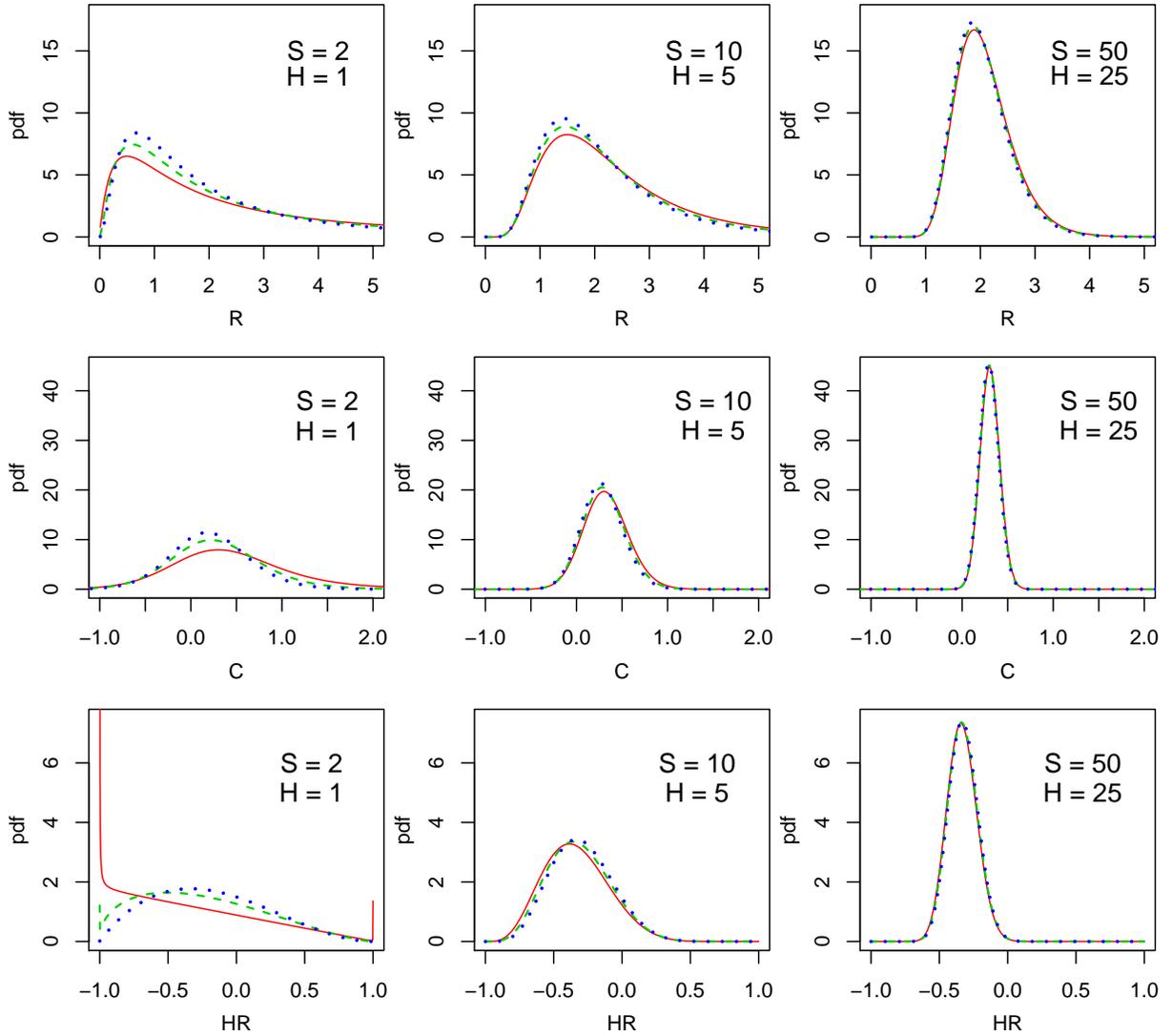}}
\caption{The posterior distributions of hardness ratios calculated
    for $\R$ (top row), $\C$ (middle row), and $\HR$ (bottom row),
    with different source intensities (low at left and high at right)
    and prior distributions. The
    different curves in each figure represent the posterior
    probability distributions computed with different
    non-informative prior distribution indices, i.e., $\phi=0^+$ (solid),
    $\phi=1/2$ (dashed), and $\phi=1$ (dotted). At higher counts
    (large Poisson intensities; right column of figures), the
    posterior distributions tends to be symmetric and the effect of
    the prior distributions is minimal, i.e., the posterior
    distributions with different indices are nearly
    the same.  At small counts (small Poisson intensities; the left
    column of figures), the non-symmetric shape of the posterior
    distribution for each hardness ratio becomes clear as does
    the effect of the choice of a non-informative prior distribution.}
\label{park:fig:dens}
\end{figure}

The posterior distribution of any hardness ratio becomes more
symmetric as both soft and hard source intensities increase.  Regardless of
the size of the intensities, however, the color has the most
symmetric posterior distribution among the popular definitions of a
hardness ratio. Figure~\ref{park:fig:dens} illustrates the effect of
the magnitude of the source intensities on the symmetry of the posterior
distribution of each hardness ratio; the posterior distribution 
of $\C$ is confirmed to have the most symmetric posterior distribution. 
In the figure, we fix $\R=2$ and the soft and hard
intensities are determined by beginning with $\lam_S=2$ and $\lam_H=1$ 
(in units of counts~(source~area)$^{-1}$)
and increasing the intensities by a factor of 5 in each subsequent 
column. We assume no background contamination in each simulation.

Figure~\ref{park:fig:dens} also illustrates the effect of the use of
different indices for the non-informative prior distribution on the
resulting posterior distribution of each hardness ratio.
In the case of $\R$, the posterior distribution becomes diffuse and more
skewed toward $0$ as the value of $\phi$ decreases; as expected, this
is also the case for $\C$.  And in the case of $\HR$, the posterior
distribution has more mass near the boundaries (i.e., $\pm1$) as the value of
$\phi$ decreases, thereby exhibiting a U-shaped density in the case of
very faint sources.  As the intensities increase, however,
the choice of a prior distribution has less effect on the posterior
distribution of any of the hardness ratios.

Finally, as pointed out in \S\ref{park:sec:verify}, for $\R$ and $\C$,
the mode of the posterior probability distribution is a robust estimator,
whereas for $\HR$, the mean of the distribution is a better choice. (Notice
that the posterior mode of $\HR$ in the low count scenario is $-1$ when 
$\phi=0^+$ in Figure~\ref{park:fig:dens}.)

\subsection{Advantages}
\label{park:sec:advan}

A significant improvement that our method provides over the classical
way of computing hardness ratios is that we use the correct Poisson
distribution throughout and do not make any limiting (high counts)
Gaussian assumptions.  Thus, while the classical method works well
only with high counts data and fails to give reliable results with low
counts, our method is valid in all regimes.  Indeed, because the
observed counts are non-negative integers, it is not appropriate to
model low counts with Gaussian distributions which are defined on all
the real numbers.

Because our methods are based on Poisson assumptions in a fully model
based statistical approach, we need not rely on plug-in estimates of
the hardness ratio. Instead we can evaluate the full posterior
probability distribution of the hardness ratio, which provides
reliable estimates and correct confidence limits even when either or
both soft and hard counts are very low.  In particular, our method
is not limited to ``detectable'' counts, and can properly calculate
upper and lower bounds in cases where the source is detected in only
one passband.  For high counts, our method gives more precise error
bars for the hardness ratios than the classical method (as defined by
the coverage; see Table~\ref{park:tbl:coverage} and
Appendix~\ref{sec:appA}), although both methods yield similar results.

Further, {\sl a priori} information can be embedded into our model and
can be updated by data, thereby producing more accurate results as we
collect more data.

\subsection{Limitations}
\label{park:sec:limit}

Unlike the quantile-width method (Hong et al.\ 2004), the
calculation of hardness ratios is limited by necessity to
pre-selected non-overlapping passbands.  In general, we must use
different band definitions in order to achieve the maximum
discriminating power for soft and hard sources.  {\sl A priori}, with
no knowledge of the character of the source populations, there is no
way to ensure that the chosen passbands are the most efficient for
splitting the spectrum and obtaining maximum leverage to distinguish
spectral model parameters.  However, this restriction allows
for a more uniform analysis and comparison of results across
different instruments and telescopes.

Because the Bayesian method does not allow for a simple analytic
solution similar to standard error propagation in the Gaussian case,
the computational methods used to determine the probability
distributions become important.  We have implemented a Monte Carlo
integration method (the Gibbs sampler, see
Appendix~\ref{park:sec:gibbs}) and a method based on numerical
integration (quadrature, see Appendix~\ref{park:sec:quad}).  The Gibbs
sampler is based on Monte Carlo simulation and hence causes our
estimates to have simulation errors in addition to their true
variability; to reduce the levels of simulation errors, the Markov
chain must be run with large enough iterations. On the other hand, the
quadrature precisely computes the posterior distribution as long as
the number of bins is large enough; however, its computation becomes
expensive as the counts become large because of the binomial expansion
in Equation~\ref{park:eq:binom-expan}. In general, the Gibbs sampler is
much faster than the method based on quadrature, but care must be
taken to ensure that the number of iterations is sufficient to
determine the posterior mode and the required posterior interval
with good precision. We recommend using the Gibbs sampler for high counts
and the quadrature for low counts.

\section{Summary}
\label{park:sec:conclusion}

We have developed a method to compute hardness ratios of counts
in non-overlapping passbands that is fully Bayesian and properly
takes into account
\begin{enumerate}
\item the Poissonian nature of the data,
\item background counts, allowing for differences in collecting
areas, effective areas, and exposure times between the source and
background regions, and
\item both informative and non-informative prior information.
\end{enumerate}
The algorithm is implemented in two forms: a Monte Carlo based
Gibbs sampler that samples from the posterior probability
distributions of the counts intensities, and another that
numerically integrates the joint probability distribution of
the counts intensities using quadrature.

We carry out comparison tests between our method and the
classical method, and show that the Bayesian method works
even in the low counts regime where the classical method
fails, and matches the results of the classical method at
high counts where a Gaussian approximation of the errors
is appropriate.

We apply the method to real world test cases of
\begin{enumerate}
\item determining candidate quiescent Low-Mass X-ray
Binaries in the Terzan-5 cluster, where we show that
it can be applied even in cases where the classical
method is entirely inapplicable, such as those
where the source is detected in only one passband, and
\item tracking the time evolution of a flare on the
dMe star Ross\,154, where we demonstrate the flexibility
of the method in being able to work around the conflicting
limitations imposed by time bin sizes and time resolution.
\end{enumerate}
Additional applications of these methods include spectral
classification of the X-ray sources detected in deep surveys and other
galaxies, and spectral variability of AGNs and X-ray binaries.
The physical properties of the source influence
the source intensities $\lam_S$ and $\lam_H$; given a source
spectral model, the two quantities are related.  Thus, variations
in the hardness ratios reflect changes in the source properties
(e.g., sources with different temperatures have different colors).
A relation between source properties, physical model parameters and
hardness ratio is a natural extension of this work and will be
presented in the future.

\acknowledgments
We thank Yue Wu for assistance with code development, and Dong-Woo Kim and Peter
Edmonds for useful discussions.
The authors gratefully acknowledge funding for this project partially
provided by NSF grants DMS-01-04129, DMS-04-38240, and DMS-04-06085
and by NASA Contracts NAS8-39073 and NAS8-03060 (CXC), and NASA grant
NAG5-13056.  This work is a product of joint work with the
California-Harvard astrostatistics collaboration (CHASC) whose members
are TP, VK, AS, DvD, and AZ, and J.\ Chiang, A.\ Connors,
D.\ Esch, P.\ Freeman, H.\ Kang, X.-L.\ Meng, J.\ Scargle, E.\ Sourlas,
and Y.\ Yu.

\appendix

\section{Computing the Hardness Ratios \label{sec:appA}}

\subsection{Monte Carlo Integration: Gibbs Sampler}
\label{park:sec:gibbs}

Monte Carlo methods can be used in Bayesian statistical inference if
representative values of the model parameters can be simulated from
the posterior distribution. Given a sufficient simulation size, these
simulated values can be used to summarize and describe the posterior
distribution and hence the Bayesian statistical inference. For
example, the posterior mean can be computed by averaging these
simulated values and posterior intervals are computed using quantiles
of the the simulated values. Markov chain Monte Carlo (MCMC) methods
(Tanner \& Wong 1987) accomplish Monte Carlo simulation by
constructing a Markov chain with stationary distribution equal to the
target posterior distribution. The Markov chain is then run, and, upon
convergence delivers the desired Monte Carlo simulation of the
posterior distribution. The Gibbs sampler (Geman \& Geman 1984) is a
special case of MCMC that constructs the Markov chain through
alternating conditional simulation of a subset of the model
parameters conditioning on the others. A more detailed description of
these methods can be found in van Dyk (2003) and the Appendix of van
Dyk et al. (2001).

In order to construct a Gibbs sampler, we embed the statistical model
described in \S\ref{park:sec:poisson} into a larger model.  Since the
observed counts are the sum of the of source counts ($\eta$) and the
background counts ($\beta$), we write $S=\eta_S+\beta_S$ and
$H=\eta_H+\beta_H$.  That is, the source counts are modeled as
independent Poisson random variables,
\begin{eqnarray}
  \eta_S \sim \Pois(\eff_S \cdot \lam_S) &\mbox{and}& 
  \eta_H \sim \Pois(\eff_H \cdot \lam_H), \label{park:eq:eta}
\end{eqnarray}
and the background counts in the source exposure area as independent 
Poisson random variables,
\begin{eqnarray}
  \:\be_S \sim \Pois(\eff_S \cdot \xi_S) &\mbox{and}& 
  \be_H \sim \Pois(\eff_H \cdot \xi_H), \label{park:eq:beta}
\end{eqnarray}
where $\lam$ and $\xi$ denote the expected source and background
counts in the source region. This implies Equation~\ref{park:eq:SH},
where the detected photons in the source region are a convolution 
of the source and background counts, i.e., $S=\be_S+\eta_S$ and 
$H=\be_H+\eta_H$.

Using the method of data augmentation, we treat the background counts
in the source exposure area ($\be_S$ and $\be_H$) as missing data; the
source counts ($\eta_S$ and $\eta_H$) are fully determined when
$\be_S$ and $\be_H$ are known. Intuitively, it is straightforward to
estimate the Poisson intensities if detected photons are split into
the source and background counts. Based on this setting, the Gibbs
sampler produces Monte Carlo draws of $\lam_S$ and $\lam_H$ along with
stochastically imputing the missing data ($\be_S$ and $\be_H$). The
conditional independence of $\lam_S$ and $\lam_H$ implies that a Monte
Carlo simulation of each hardness ratio can be obtained by simply
transforming that of $\lam_S$ and $\lam_H$. Due to the conditional
independence, both soft and hard bands also have exactly the same
sampling steps, and thus we illustrate the Gibbs sampler only for the
soft band. First, the joint posterior distribution of $\lam_S$,
$\xi_S$, and $\be_S$ is given by
\begin{eqnarray}
  p(\lam_S,\xi_S,\be_S|S,B_S) 
    &\propto& p(S|\lam_S,\be_S)p(B_S|\xi_S)p(\be_S|\xi_S)
	p(\lam_S)p(\xi_S) \nonumber\\
    &\propto& \frac{1}{(S-\be_S)!\be_S!}
	      \lam^{S-\be_S+\psi_{S_1}-1}\xi^{B+\be_S+\psi_{S_3}-1}
	      e^{-(\eff_S+\psi_{S_2})\lam_S-(\eff_S + {r}\cdot\eff_S +\psi_{S_4})\xi_S}.\nonumber\\
    \label{park:eq:gibbs-joint}
\end{eqnarray}
That is, conditional on the total soft counts ($S$), the unobserved
background counts in the source exposure area ($\be_S$) follows a
binomial distribution: Given the current iterates of the parameters,
$\lam_S^{(t)}$ and $\xi_S^{(t)}$, \step~1 is given by
\begin{description}
  \item[\step~1 :] Draw $\be_S^{(t+1)}$ from 
	$\be_S|\big(\lam_S^{(t)},\xi_S^{(t)},S,B_S\big) \sim 
	\Binom\l(S,\:\ds\frac{\xi_S^{(t)}}{\lam_S^{(t)}+\xi_S^{(t)}}\r)$,
\end{description}
for $t=1,\dots,T$. Next, \steps~2 and~3 draw the source and 
background intensities from the $\gamma$-distributions. 
In particular, \steps~2 and~3 find the next iterates of the 
intensities using
\begin{description}
  \item[\step~2 :] Draw $\lam_S^{(t+1)}$ from 
	$\lam_S|\big(\xi_S^{(t)},\be_S^{(t+1)},S,B_S\big)\sim 
	\Gam\big(S-\be_S^{(t+1)}+\psi_{S_1},\:\eff_S +\psi_{S_2}\big)$
  \mbox{ and}
  \item[\step~3 :] Draw $\xi_S^{(t+1)}\,$ from 
	$\xi_S|\big(\lam_S^{(t+1)},\be_S^{(t+1)},S,B_S\big)\sim 
	\Gam\big(B_S+\be_S^{(t+1)}+\psi_{S_3},\:\eff_S + {r}\cdot\eff_S +\psi_{S_4}\big)$,
\end{description}
for $t=1,\dots,T$. To accomplish one Gibbs iteration, these steps are
implemented for the soft and hard bands. After iterating the Gibbs
sampler $T$ times, we collect a posterior sample
$\{\lam_S^{(t)},\,\lam_H^{(t)},\:t=t_0+1,\dots,T\}$ for a sufficiently
long burn-in period\footnote{
The term ``burn-in'' refers to the
number of iterations necessary for the Markov chain to converge
and begin to sample from the posterior probability distribution.}
$t_0$. The analytical calculation to determine
burn-in is far from computationally feasible in most
situations. However, visual inspection of plots of the Monte Carlo
output is commonly used for determining burn-in. More formal tools for
determining $t_0$, called convergence diagnostics, have been proposed;
for a recent review, see Cowles \& Carlin (1996).  Under a monotone
transformation (see Equations~\ref{e:postR},\ref{e:postC},\ref{e:postHR})
of the posterior samples, $(T-t_0)$ Monte Carlo draws
for each hardness ratio are obtained, which enables us to find its
point estimates and the corresponding error bar; see
\S\ref{park:sec:stats} for details.

\subsection{Numerical Integration: Gaussian Quadrature}
\label{park:sec:quad}

Instead of introducing the missing data, the Bayes' theorem can
analytically compute the joint posterior distribution of $\lam_S$ and
$\lam_H$, based on which we obtain the posterior distribution of each
hardness ratio through quadrature. Because the
models for the hard and soft bands are independent and symmetric, we
illustrate the computation only for the soft intensity $\lam_S$.
First, the joint posterior distribution of $\lam_S$ and $\xi_S$
in Equation~\ref{park:eq:Bayes} is written out as
\begin{eqnarray}
  p(\lam_S,\xi_S|S,B_S) 
  &=& \frac{p(\lam_S)p(\xi_S)p(S|\lam_S,\xi_S)p(B_S|\xi_S)}
	{\int_0^{\infty}\!\!\int_0^{\infty}p(\lam_S)p(\xi_S)p(S|\lam_S,\xi_S)
	p(B_S|\xi_S)d\xi_S\, d\lam_S}\nonumber\\
  &=& \frac{(\lam_S+\xi_S)^{S}\lam_S^{\psi_{S_1}-1}\xi_S^{B_S+\psi_{S_3}-1}
	e^{-(\eff_S+\psi_{S_2})\lam_S}e^{-(\eff_S +{r}\cdot\eff_S +\psi_{S_4})\xi_S}}
	{\int_0^{\infty}\!\!\int_0^{\infty}(\lam_S+\xi_S)^{S}
	\lam_S^{\psi_{S_1}-1}\xi_S^{B_S+\psi_{S_3}-1}
	e^{-(\eff_S +\psi_{S_2})\lam_S}e^{-(\eff_S +{r}\cdot\eff_S +\psi_{S_4})\xi_S}d\xi_Sd\lam_S}.\nonumber\\
	\label{park:eq:quad-joint}
\end{eqnarray}
Then, the binomial expansion to $(\lam_S+\xi_S)^{S}$, i.e.,
\begin{eqnarray}
  (\lam_S+\xi_S)^{S}
  =\sum_{j=0}^S \frac{\Gamma(S+1)}{\Gamma(j+1)\Gamma(S-j+1)}
	 \lam_S^j\,\xi_S^{S-j},
	\label{park:eq:binom-expan}		
\end{eqnarray}
enables us to analytically integrate $\xi_S$ out of the joint
distribution in Equation~\ref{park:eq:quad-joint}; and obtain an analytical
solution to the marginal posterior distribution of $\lam_S$.
Therefore, using the
binomial expansion, Equation~\ref{park:eq:quad-joint} becomes
\begin{eqnarray}
  p(\lam_S,\xi_S|S,B_S) = 
	\frac{\ds\sum_{j=0}^S \frac{\Gamma(S+1)}{\Gamma(j+1)\Gamma(S-j+1)}
	 \lam_S^{j+\psi_{S_1}-1}\xi_S^{S-j+B_S+\psi_{S_3}-1}
	e^{-(\eff_S +\psi_{S_2})\lam_S}e^{-(\eff_S+{r}\cdot\eff_S+\psi_{S_4})\xi_S}}
	{\ds\sum_{j=0}^S \frac{\Gamma(S+1)}{\Gamma(j+1)\Gamma(S-j+1)}
	\cdot\frac{\Gamma(S-j+B_S+\psi_{S_3})} {(\eff_S+{r}\cdot\eff_S+\psi_{S_4})^{S-j+B_S+\psi_{S_3}}} 
	\cdot\frac{\Gamma(j+\psi_{S_1})}{(\eff_S+\psi_{S_2})^{j+\psi_{S_1}}}}
	\label{park:eq:quad-joint2}
\end{eqnarray}
and then $\xi_S$ is integrated out of Equation~\ref{park:eq:quad-joint2}, i.e.,
\begin{eqnarray}
  p(\lam_S|S,B_S) = 
	\frac{\ds\sum_{j=0}^S \frac{1}{\Gamma(j+1)\Gamma(S-j+1)}
	\cdot\frac{\Gamma(S-j+B_S+\psi_{S_3})}{(\eff_S+{r}\cdot\eff_S+\psi_{S_4})^{S-j+B_S+\psi_{S_3}}}
	\lam_S^{j+\psi_{S_1}-1}e^{-(\eff_S+\psi_{S_2})\lam_S}}
	{\ds\sum_{j=0}^S \frac{1}{\Gamma(j+1)\Gamma(S-j+1)}
	\cdot\frac{\Gamma(S-j+B_S+\psi_{S_3})}{(\eff_S+{r}\cdot\eff_S+\psi_{S_4})^{S-j+B_S+\psi_{S_3}}}
	\cdot\frac{\Gamma(j+\psi_{S_1})}{(\eff_S+\psi_{S_2})^{j+\psi_{S_1}}}}.
	\label{park:eq:margin}
\end{eqnarray}
Here the prior independence of $\lam_S$ and $\lam_H$ implies that the
joint posterior distribution of these two intensities is
decomposed into a product of their marginal posterior distributions, i.e.,
\begin{eqnarray}
  p(\lam_S,\lam_H|S,H,B_S,B_H)=p(\lam_S|S,B_S)p(\lam_H|H,B_H).
  \label{park:eq:joint}
\end{eqnarray}
This assumption of independence between $\lam_S$ and $\lam_H$ can be
relaxed, and hierarchical modeling for sources to determine
clustering properties and spectral parameters can be devised
(in preparation).

Using the joint posterior distribution in Equation~\ref{park:eq:joint}, we
compute the analytical solution to the posterior distribution of each
hardness ratio as follows:
\begin{enumerate}
\item the posterior distribution of $\R$ is obtained after integrating 
$\lam_H$ out of
\begin{eqnarray*}
  && p(\R,\lam_H|S,H,B_S,B_H)\,d\R\,d\lam_H \\
  && =\ p(\lam_S,\lam_H|S,H,B_S,B_H)\bigg|\frac{\partial(\lam_S,\lam_H)}
	{\partial(\R,\lam_H)}\bigg|\,d\lam_S\,d{\lam_H}\\
  && =\ p(\R\lam_H,\lam_H|S,H,B_S,B_H)\lam_H\,d\R\,d{\lam_H},
\end{eqnarray*}
\item the posterior distribution of $\C$ is obtained after integrating
$\lam_H$ out of
\begin{eqnarray*}
  && p(\C,\lam_H|S,H,B_S,B_H)\,d\C\,d\lam_H \\ 
  && =\ p(\lam_S,\lam_H|S,H,B_S,B_H)\bigg|\frac{\partial(\lam_S,\lam_H)}
	{\partial(\C,\lam_H)}\bigg|\,d\lam_S\,d\lam_H \\
  && =\ p(10^{\C}\lam_H,\lam_H|S,H,B_S,B_H)
	10^{\C}\ln(10)\lam_H\,d\C\,d{\lam_H}, \mbox{ and}
\end{eqnarray*}
\item the posterior distribution of $\HR$ is obtained after integrating
$\omega=\lam_S+\lam_H$ out of
\begin{eqnarray*}
  && p(\HR,\omega|S,H,B_S,B_H)\,d\HR\,d\omega \\
  && =\ p(\lam_S,\lam_H|S,H,B_S,B_H)\bigg|\frac{\partial(\lam_S,\lam_H)}
	{\partial(\HR,\omega)}\bigg|\,d\lam_S\,d{\lam_H}\\
  && =\ p\bigg(\frac{(1-\HR)\omega}{2},\frac{(1+\HR)\omega}{2}
	\Big|S,H,B_S,B_H\bigg)\frac{\omega}{2}\,d\HR\,d{\omega}.
\end{eqnarray*}
\end{enumerate}
To numerically integrate out a parameter in these analytical
solutions, we employ quadrature, which precisely evaluates the
integral of a real valued function (and has nothing to do with the
Gaussian counts statistic assumptions); refer to Wichura (1989) for details of the
algorithm. Due to the complex expression, the probability density for each
hardness ratio is approximated for each equally-spaced abscissas equally spaced
over the finite range of the hardness ratio. Then we summarize the
approximate posterior density in order to calculate the inferences, as
detailed in \S\ref{park:sec:stats}.

\section{Effect of Non-informative Priors}
\label{sec:appB}

In order to compare the effect of these non-informative prior
distributions, we have carried out simulations for low counts and high
counts cases to test the computed coverage against the simulation:
Table~\ref{park:tbl:idxes1} 
presents the
coverage rate and mean length of the 95\% intervals for the color,
$\C$.  In order to simplify the comparison, we assume that there is no
background contamination, and simulate 1000 realizations of source
counts for each case.  The tables are laid out as a grid of soft and
hard source counts per source area (i.e., $\lambda_S$ versus
$\lambda_H$, see Equation~\ref{park:eq:eta}).
For each $(\lambda_S,\lambda_H)$ pair,
we report the percentage of the simulations that result in 95\%
posterior intervals that actually contain $(\lambda_S,\lambda_H)$
along with the mean length of the 95\% posterior intervals from the
simulations.  This calculation is carried out for three choices of the
prior index, $\phi=0^+, 1/2, 1$, corresponding to the top, middle,
and bottom elements in each cell of the table.

\begin{table}[t]
  \begin{center}
  \caption{Legend Key for Table~\ref{park:tbl:idxes1}.}
  \medskip
  \begin{tabular}{|c|c|c|}
\cline{2-3}
\divi{} & \multicolumn{2}{c|}{Hard band source intensity,} \\
\divi{} & \multicolumn{2}{c|}{$\lambda_H$ [counts~(source area)$^{-1}$]} \\
\hline
  & Coverage rate of          & Average length of \\[-.2em]
  & intervals with $\phi=0^+$ & intervals with $\phi=0^+$ \\[.5em]
  Soft band source intensity
  & Coverage rate of          & Average length of \\[-.2em]
  $\lambda_S$ [counts~(source area)$^{-1}$]
  & intervals with $\phi=1/2$ & intervals with $\phi=1/2$ \\[.5em]
  & Coverage rate of          & Average length of \\[-.2em]
  & intervals with $\phi=1$ & intervals with $\phi=1$ \\
\hline
  \end{tabular}
  \label{park:tbl:legend}
  \end{center}
\end{table}

In these simulations, we theoretically expect that the computed 95\%
posterior intervals contain the actual value with a probability of 95\%.
Due to the Monte Carlo simulation errors, we expect most coverage rates to be
between 0.93 and 0.97 which are three standard deviations away from
0.95; a standard deviation of the coverage probability for 95\%
posterior intervals is given by $\sqrt{(0.95)(0.05)/1000}=0.0069$ under
a binomial model for the Monte Carlo simulation.

\begin{figure}[t]
\centerline{\includegraphics[width=7in,angle=0]{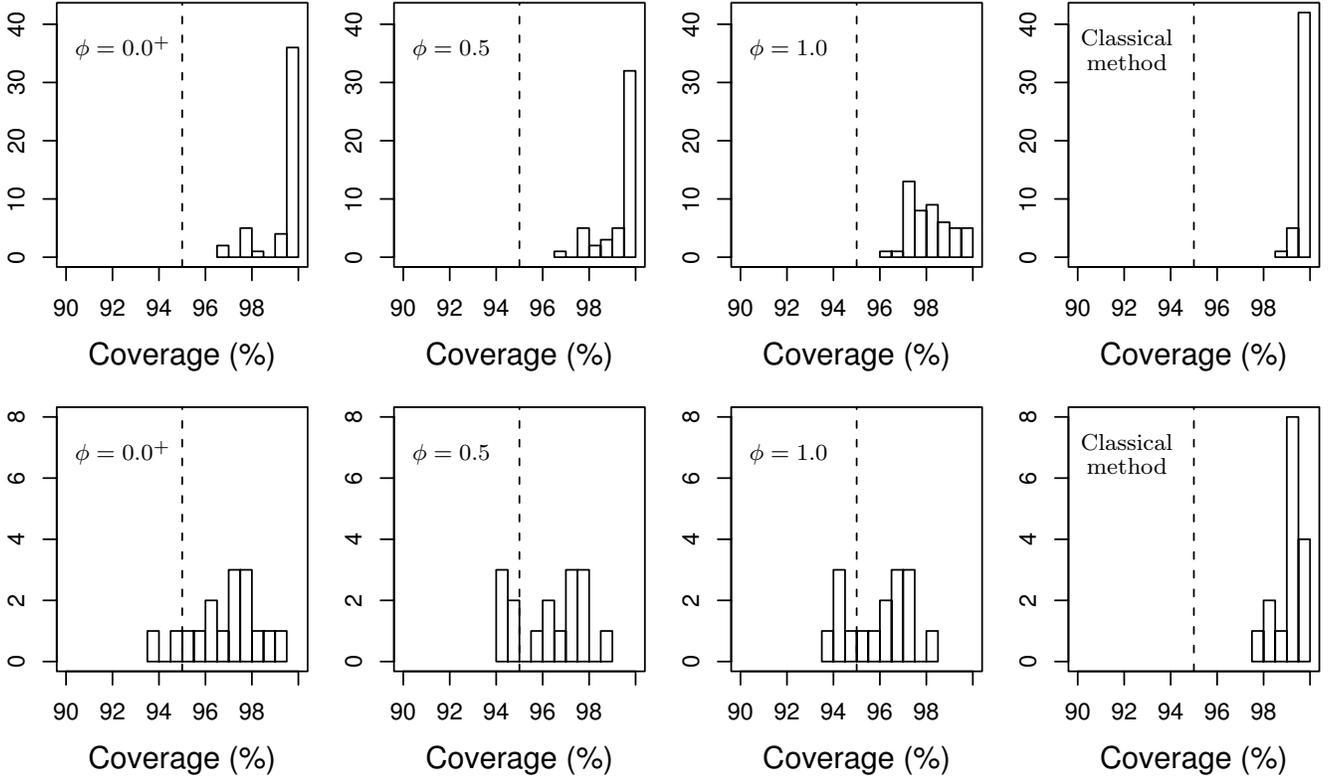}}
\caption{
    The empirical distributions of coverage rates
    with different prior distribution indices.
    The first three
    columns of the panels are constructed under different types of
    non-informative prior distributions, $\phi=0^+$ (first),
    $\phi=1/2$ (second), $\phi=1$ (third).  The fourth column 
    depicts results from the classical method, and is presented
    for reference.  (In calculating the coverage and the length of
    the intervals, We exclude simulations where $S=0$ or $H=0$ since
    $\C$ is undefined in such cases.)
    The top panels correspond to the low count scenarios
    $\lam_S=(0.5,1,2,4)$ or $\lam_H=(0.5,1,2,4)$ and the bottom panels
    correspond to the remaining cases.
    Coverage rate is the
    fraction of simulated intervals that contain the parameter under
    which the data were simulated.  In this case, we compute
    the coverage rate based on the 95\% posterior intervals for $\C$
    with different combinations of values for $\lam_S$ and $\lam_H$.
    The histograms represent the empirical distribution of the
    coverage rate. The dotted vertical lines represent the
    expected coverage rate of 95\%.
    \label{park:fig:cover} }
\end{figure}

\begin{figure}[t]
{\includegraphics[width=3.5in]{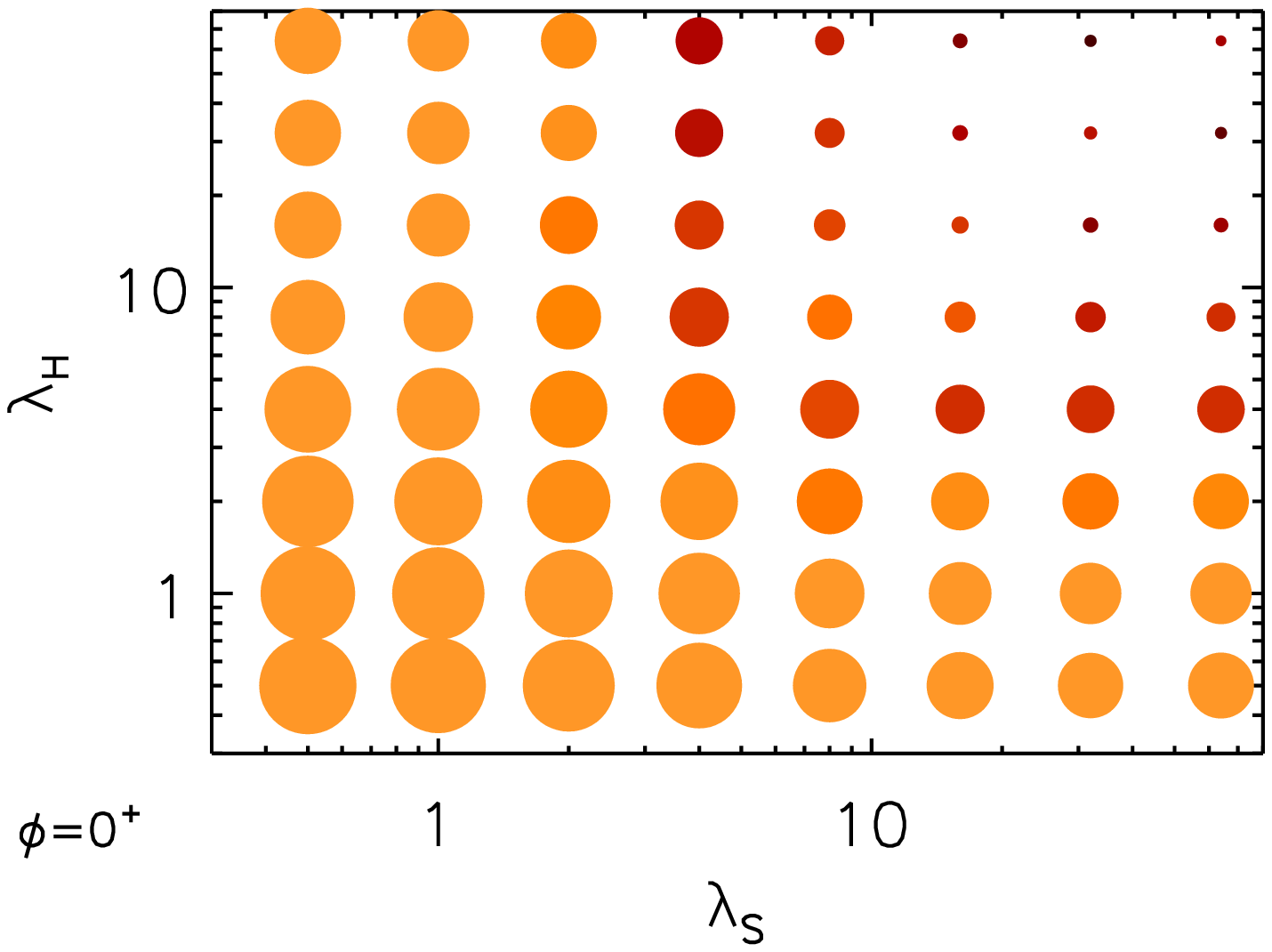}\\
\includegraphics[width=3.5in]{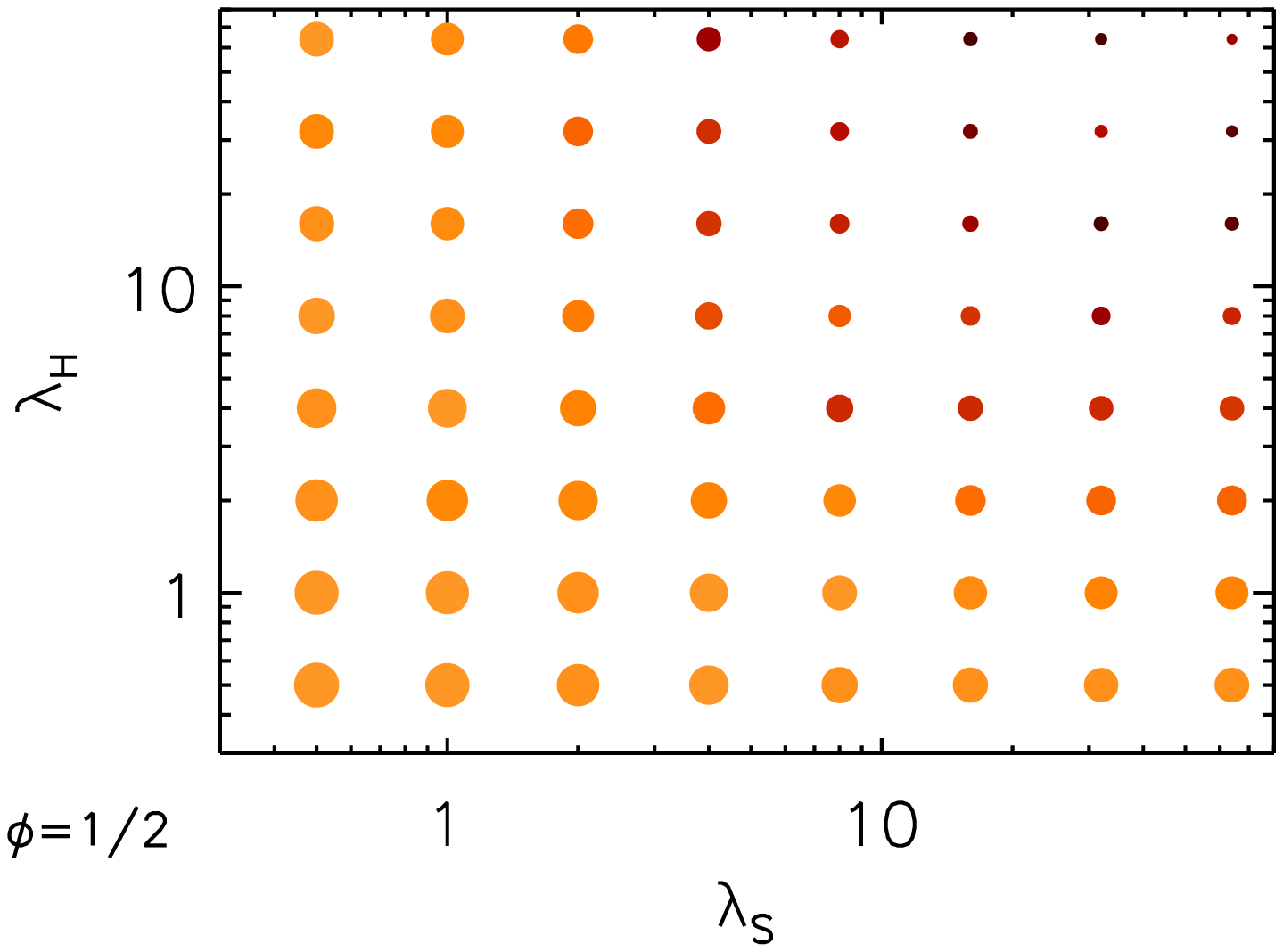}\\
\includegraphics[width=3.5in]{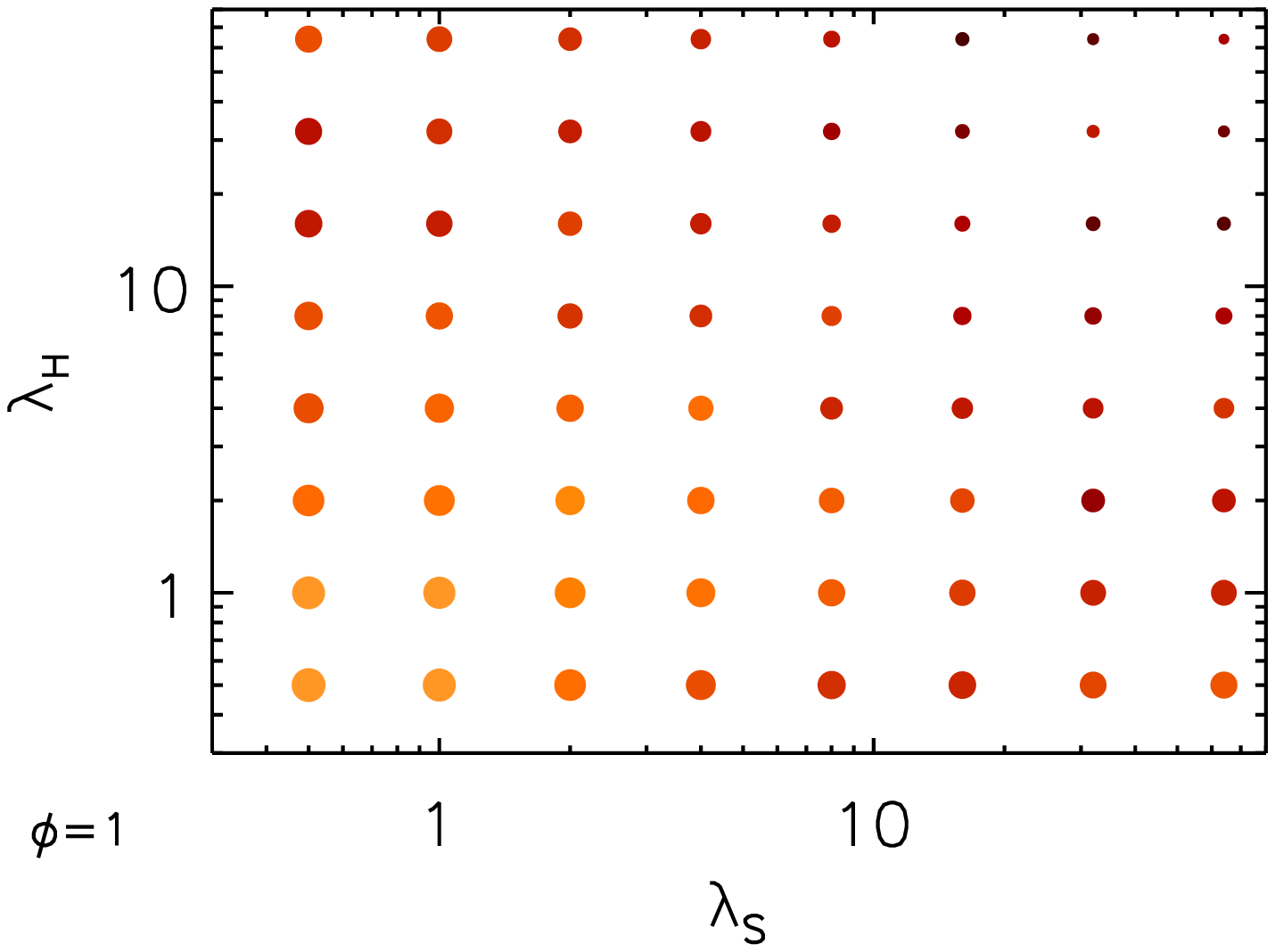}}
\caption{
Graphical representation of the data in Table~\ref{park:tbl:idxes1}.
The symbols are located on a grid corresponding to the appropriate
$\lam_S$ (absissa) and $\lam_H$ (ordinate), for $\phi=0^+$ (top),
$\phi=1/2$ (middle), and $\phi=1$ (bottom).
The shading of each symbol represents the coverage, with 100\% being
lightest and progressively getting darker as the coverage percentage
decreases.  The sizes of the symbols are scaled as the
$\sqrt{{\rm length~of~interval}}$.  Note that for small values of $\phi$,
which correspond to a prior expectation of a large dynamic range in
the source colors, the intervals are large when the counts are small
(i.e., the choice of prior distribution has a large effect), and decrease
to values similar to those at larger $\phi$ when there are more counts.
\label{fig:idxesfig}}
\end{figure}

Table~\ref{park:tbl:idxes1} presents the
coverage rate and average length of posterior intervals for small
and large magnitudes of source intensities.  The key
to this table is given in
Table~\ref{park:tbl:legend}: For each $(\lam_S,\lam_H)$ pair, the
posterior intervals are simulated using different prior distribution
indices ($\phi=0^+$, 1/2, 1) and the summary statistics -- the
coverage rate and the mean lengths of the intervals -- are
displayed from top ($\phi=0^+$) to bottom ($\phi=1$) within 
each cell.
The same information is shown in graphical form in
Figure~\ref{fig:idxesfig}.
The use of $\phi=0^+$ tends to yield
very wide posterior intervals and under-cover the true X-ray color
when the source intensities are low.  On the other hand,
the other two non-informative prior distribution indices produce
much shorter posterior intervals and maintain high coverage rates.
With high counts, however,
the choice of a non-informative prior distribution does not have a
noticeable effect on the statistical properties of the posterior
intervals.  The same results are summarized in graphical form in
Figure~\ref{park:fig:cover}, where the 95\% coverage rates are shown
for various cases.
The empirical distributions of the coverage rate computed under
either low count ($\lam_S=0.5,1,2,4$ or $\lam_H=0.5,1,2,4$) or high count
(all of the remaining) scenarios are shown, along
with a comparison with the results from the classical method
(last column).  The distinction
between the Bayesian and classical methods is very clear in this
figure.
Note that the posterior intervals
over-cover relative to our theoretical expectation, but to a lesser
degree than the classical intervals. Over-coverage is conservative and
preferred to under-coverage, but should be minimal when possible.
Considering the low count and high count scenarios,
Figure~\ref{park:fig:cover} suggests using the Jeffrey's
non-informative prior distribution (i.e., $\phi=1/2$) in general.
At low counts, the shape (and consequently the coverage) of the
posterior distribution is dominated by our prior assumptions, as
encoded in the parameter $\phi$.  The coverage rate will also be
improved when informative priors are used.

\clearpage
\pagestyle{empty}
\setlength{\voffset}{50mm}
{\rotate
\begin{table}
  \begin{center}
  \caption{Coverage of the X-ray Color ($\C$) Using the Bayesian Method with
       Different (see Table~\ref{park:tbl:legend} for key)
       Non-Informative Prior Distribution Indices
       ($\phi=0^+, 1/2, 1$).}
  \medskip
  \begin{tabular}{|c|c|r|r|r|r|r|r|r|r|r|r|r|r|r|r|r|r|}
\cline{3-18}
 \mdivi{}&\multicolumn{16}{c|}{$\lam_H$} \\
\cline{3-18}
 \mdivi{}&\mdivi{0.5}&\mdivi{1.0}&\mdivi{2.0}&\mdivi{4.0}&
     \mdivi{8.0}&\mdivi{16.0}&\mdivi{32.0}&\mdivi{64.0}\\
  \hline
  &     & 100  \%& 23.99 & 100  \%& 22.67 & 100  \%& 21.19 & 100  \%& 19.02 & 100  \%& 14.07 & 100  \%& 11.32 & 100  \%& 11.14 & 100  \%& 11.11 \\
  & 0.5 & 100  \%& 5.10  & 100  \%& 4.88  & 99.9 \%& 4.51  & 99.9 \%& 3.95  & 100  \%& 3.35  & 99.9 \%& 3.09  & 99.7 \%& 3.02  & 100  \%& 3.03 \\
  &     & 100  \%& 2.87  & 100  \%& 2.71  & 99.0 \%& 2.50  & 98.4 \%& 2.25  & 98.4 \%& 2.02  & 97.2 \%& 1.90  & 97.0 \%& 1.82  & 98.4 \%& 1.83 \\
     \cline{2-18}
  &     & 100  \%& 22.95 & 100  \%& 21.59 & 100  \%& 19.61 & 100  \%& 17.40 & 100  \%& 12.24 & 100  \%& 10.03 & 100  \%& 9.84  & 100  \%& 9.51 \\
  & 1.0 & 100  \%& 4.91  & 100  \%& 4.69  & 99.7 \%& 4.31  & 100  \%& 3.77  & 99.9 \%& 3.04  & 99.8 \%& 2.80  & 99.7 \%& 2.74  & 99.8 \%& 2.71 \\
  &     & 100  \%& 2.73  & 100  \%& 2.58  & 99.2 \%& 2.37  & 98.9 \%& 2.14  & 98.5 \%& 1.85  & 97.3 \%& 1.73  & 97.7 \%& 1.67  & 98.0 \%& 1.65 \\
     \cline{2-18}
  &     & 100  \%& 21.42 & 100  \%& 19.57 & 99.8 \%& 17.52 & 99.7 \%& 15.04 & 99.6 \%& 10.54 & 99.3 \%& 8.45  & 99.9 \%& 7.95  & 99.8 \%& 7.82 \\
  & 2.0 & 99.9 \%& 4.52  & 99.9 \%& 4.30  & 99.7 \%& 3.89  & 99.6 \%& 3.27  & 99.5 \%& 2.56  & 99.2 \%& 2.33  & 99.0 \%& 2.21  & 99.4 \%& 2.21 \\
  &     & 99.1 \%& 2.51  & 99.5 \%& 2.37  & 99.7 \%& 2.14  & 98.8 \%& 1.87  & 97.8 \%& 1.60  & 98.1 \%& 1.49  & 97.3 \%& 1.40  & 97.7 \%& 1.39 \\
     \cline{2-18}
  &     & 100  \%& 18.60 & 100  \%& 16.80 & 99.9 \%& 15.18 & 99.2 \%& 13.11 & 97.9 \%& 8.87  & 97.9 \%& 6.06  & 97.0 \%& 5.85  & 96.8 \%& 5.63 \\
  & 4.0 & 100  \%& 3.92  & 100  \%& 3.68  & 99.6 \%& 3.30  & 99.2 \%& 2.64  & 98.5 \%& 1.89  & 98.0 \%& 1.60  & 97.9 \%& 1.53  & 96.6 \%& 1.49 \\
  &     & 98.4 \%& 2.24  & 99.2 \%& 2.09  & 99.0 \%& 1.87  & 99.1 \%& 1.59  & 97.7 \%& 1.29  & 97.3 \%& 1.14  & 97.1 \%& 1.08  & 97.4 \%& 1.04 \\
     \cline{2-18}\up{$\lam_S$}
  &     & 100  \%& 13.66 & 100  \%& 12.28 & 99.3 \%& 10.93 & 98.3 \%& 8.74  & 99.2 \%& 5.13  & 98.2 \%& 2.50  & 97.8 \%& 2.26  & 97.4 \%& 2.14 \\
  & 8.0 & 99.9 \%& 3.29  & 100  \%& 3.04  & 99.7 \%& 2.64  & 97.8 \%& 1.86  & 98.8 \%& 1.25  & 97.6 \%& 0.96  & 97.2 \%& 0.86  & 97.4 \%& 0.83 \\
  &     & 97.7 \%& 1.99  & 98.7 \%& 1.85  & 98.7 \%& 1.63  & 97.5 \%& 1.28  & 98.1 \%& 1.01  & 97.3 \%& 0.83  & 96.3 \%& 0.75  & 97.1 \%& 0.71 \\
     \cline{2-18}
  &     & 100  \%& 11.37 & 100  \%& 9.96  & 99.8 \%& 8.49  & 97.7 \%& 6.11  & 98.6 \%& 2.41  & 97.9 \%& 0.73  & 96.7 \%& 0.60  & 95.5 \%& 0.54 \\
  &16.0 & 99.9 \%& 3.10  & 99.8 \%& 2.79  & 99.2 \%& 2.34  & 97.8 \%& 1.60  & 98.0 \%& 0.95  & 96.6 \%& 0.65  & 95.6 \%& 0.55  & 94.4 \%& 0.50 \\
  &     & 97.5 \%& 1.91  & 98.0 \%& 1.73  & 98.2 \%& 1.49  & 97.2 \%& 1.14  & 96.7 \%& 0.83  & 96.6 \%& 0.63  & 95.3 \%& 0.54  & 93.7 \%& 0.49 \\
     \cline{2-18}
  &     & 100  \%& 10.82 & 100  \%& 9.56  & 99.3 \%& 8.00  & 97.7 \%& 5.74  & 97.3 \%& 2.33  & 95.7 \%& 0.59  & 97.1 \%& 0.43  & 93.8 \%& 0.37 \\
  &32.0 & 99.9 \%& 2.98  & 99.6 \%& 2.71  & 99.0 \%& 2.21  & 97.8 \%& 1.51  & 96.5 \%& 0.88  & 94.4 \%& 0.55  & 97.2 \%& 0.43  & 94.3 \%& 0.36 \\
  &     & 98.2 \%& 1.80  & 97.4 \%& 1.65  & 96.0 \%& 1.40  & 97.1 \%& 1.06  & 96.0 \%& 0.75  & 94.5 \%& 0.54  & 97.2 \%& 0.42  & 94.5 \%& 0.36 \\
     \cline{2-18}
  &     & 100  \%& 10.99 & 100  \%& 9.52  & 99.7 \%& 7.84  & 97.7 \%& 5.69  & 97.7 \%& 2.11  & 96.3 \%& 0.53  & 94.6 \%& 0.36  & 96.5 \%& 0.29 \\
  &64.0 & 99.9 \%& 3.02  & 99.6 \%& 2.72  & 99.0 \%& 2.23  & 98.1 \%& 1.52  & 97.7 \%& 0.83  & 94.8 \%& 0.50  & 94.8 \%& 0.36  & 96.5 \%& 0.29 \\
  &     & 98.5 \%& 1.82  & 97.4 \%& 1.66  & 97.1 \%& 1.40  & 97.8 \%& 1.05  & 96.6 \%& 0.71  & 94.2 \%& 0.49  & 94.9 \%& 0.36  & 96.5 \%& 0.28 \\
  \hline
  \end{tabular}
  \label{park:tbl:idxes1}
  \end{center}
\end{table}
}
\clearpage

\setlength{\voffset}{0mm}
\pagestyle{plaintop}

\clearpage

\clearpage




\begin{references}

\reference{} Alexander, D.M., Brandt, W.N., Hornschemeier, A.E., Garmire, G.P., Schneider, D.P., Bauer, F.E., \& Griffiths, R.E., 2001, AJ, 122, 2156 
\reference{} Babu, G.J., \& Feigelson, E.D., 1997, {\sl Statistical Challenges in Modern Astronomy}, Springer-Verlag:New York
\reference{} Bradt, H., Mayer, W., Buff, J., Clark, G.W., Doxsey, R., Hearn, D., Jernigan, G., Joss, P.C., Laufer, B., Lewin, W., Li, F., Matilsky, T., McClintock, J., Primini, F., Rappaport, S., \& Schnopper, H., 1976, ApJ, 204, L67
\reference{} Brandt, W.N., Hornschemeier, A.E., Alexander, D.M., Garmire, G.P., Schneider, D.P., Broos, P.S., Townsley, L.K., Bautz, M.W., Feigelson, E.D., \& Griffiths, R.E., 2001a, AJ, 122, 1 
\reference{} Brandt, W.N., Alexander, D.M., Hornschemeier, A.E., Garmire, G.P., Schneider, D.P., Barger, A.J., Bauer, F.E., Broos, P.S., Cowie, L.L., Townsley, L.K., Burrows, D.N., Chartas, G., Feigelson, E.D., Griffiths, R.E., Nousek, J.A., \& Sargent, W.L.W., 2001b, AJ, 122, 2810 
\reference{} Brandt, W.N., \& Hasinger, G., 2005, ARAA, 43, 827
\reference{} Brown, E.F., Bildsten, L., \& Rutledge, R.E., 1998, ApJ, 504, 95
\reference{} Campana, S., Colpi, M., Mereghetti, S., Stella, L., \& Tavani, M., 1998, A\&ARv, 8, 279
\reference{} Casella, G. \& Berger, R.L., 2002, Statistical Inference, 2nd edition. Duxbury.
\reference{} Cowles, M.K., \& Carlin, B.P., 1996, J.\ Am.\ Stat.\ Assn., 91, 883
\reference{} Esch, D.N., 2003, Ph.D.\ Thesis, Department of Statistics, Harvard University
\reference{} Feigelson, E.D., \& Babu, G.J., 1992, {\sl Statistical Challenges in Modern Astronomy}, Springer-Verlag:Berlin Heidelberg New York
\reference{} Feigelson, E.D., \& Babu, G.J., 2003, {\sl Statistical challenges in astronomy. Third Statistical Challenges in Modern Astronomy (SCMA III) Conference}, University Park, PA, USA, July 18 - 21 2001, Springer:New York
\reference{} Gehrels, N., 1986, ApJ, 303, 336
\reference{} Geman, S., \& Geman, D., 1984, IEEE, 6, 721
\reference{} Giacconi, R., Rosati, P., Tozzi, P., Nonino, M., Hasinger, G., Norman, C., Bergeron, J., Borgani, S., Gilli, R., Gilmozzi, R., \& Zheng, W., 2001, ApJ, 551, 624
\reference{} Gregory, P.C., \& Loredo, T.J., 1992, ApJ, 398, 146
\reference{} Hasinger, G., \& van der Klis, M., 1989, A\&A, 225, 79
\reference{} Heinke, C.O., Grindlay, J.E., Edmonds, P.D., Lloyd, D.A., Murray, S.S., Cohn, H.N., \& Lugger, P.M., 2003, ApJ, 598, 501
\reference{} Hong, J.S., Schlegel, E.M., \& Grindlay, J.E., 2004, ApJ, 614, 508
\reference{} Gillessen, S., \& Harney, H.L., 2005, A\&A, 430, 355
\reference{} Jansen, F., Lumb, D., Altieri, B., Clavel, J., Ehle, M., Erd, C., Gabriel, C., Guainazzi, M., Gondoin, P., Much, R., Munoz, R., Santos, M., Schartel, N., Texier, D., \& Vacanti, G., 2001, A\&A, 365, L1 
\reference{} Kashyap, V.L., Drake, J.J., G\"{u}del, M., \& Audard, M., 2002, ApJ, 580, 1118
\reference{} Kashyap, V.L., \& Drake, J.J., 1998, ApJ, 503, 450
\reference{} Kim, D.-W., Barkhouse, W.A., Colmenero, E.R., Green, P.J., Kim, M., Mossman, A., Schlegel, E., Silverman, J.D., Aldcroft, T., Ivezic, Z., Anderson, C., Kashyap. V., Tananbaum, H., \& Wilkes, B.J., 2005, submitted to ApJ
\reference{} Kim, D.-W., Fabbiano, G., \& Trinchieri, G., 1992, ApJ, 393, 134
\reference{} Pallavicini, R., Tagliaferri, G., \& Stella, L., 1990, A\&A, 228, 403
\reference{} Protassov, R., van Dyk, D.A., Connors, A., Kashyap, V.L., \& Siemiginowska, A., 2002, ApJ, 571, 545
\reference{} Reale, F., Betta, R., Peres, G., Serio, S., \& McTiernan, J., 1997, A\&A, 325, 782
\reference{} Reale, F., G\"{u}del, M., Peres, G., \& Audard, M., 2004, A\&A, 416, 733
\reference{} Reale, F., Serio, S., \& Peres, G., 1993, A\&A, 272, 486
\reference{} Rosner, R., Golub, L., \& Vaiana, G.S., 1985, ARAA, 23, 413
\reference{} Schmitt, J.H.M.M., \& Favata, F., 1999, Nature, 401, 44
\reference{} Serio, S., Reale, F., Jakimiec, J., Sylwester, B., \& Sylwester, J. 1991, A\&A, 241, 197
\reference{} Silverman, J.D., Green, P.J., Barkhouse, W.A., Kim, D.-W., Aldcroft, T.L., Cameron, R.A., Wilkes, B.J., Mossman, A., Ghosh, H., Tananbaum, H., Smith, M.G., Smith, R.C., Smith, P.S., Foltz, C., Wik, D., \& Jannuzi, B.T., 2005, ApJ, 618, 123 
\reference{} Tanner, M.A. \& Wong, W.H., 1987, J.\ Am.\ Stat.\ Assn., 82, 528
\reference{} Tuohy, I.R., Garmire, G.P., Lamb, F.K., \& Mason, K.O., 1978, ApJ, 226, L17
\reference{} Vaiana, G.S., Cassinelli, J.P., Fabbiano, G., Giacconi, R., Golub, L., Gorenstein, P., Haisch, B.M., Harnden, F.R.,Jr., Johnson, H.M., Linsky, J.L., Maxson, C.W., Mewe, R., Rosner, R., Seward, F., Topka, K., \& Zwaan, C., 1981, ApJ, 245, 163
\reference{} van den Oord, G.H.J., \& Mewe, R., 1989, A\&A, 213
\reference{} van Dyk, D.A., 2003, in {\it Statistical challenges in astronomy. Third Statistical Challenges in Modern Astronomy (SCMA III) Conference}, University Park, PA, USA, July 2001, Eds.\ E.D.Feigelson, G.J.Babu, New York: Springer, p41
\reference{} van Dyk, D.A., Connors, A., Kashyap, V.L., \& Siemiginowska, A., 2001, ApJ 548, 224
\reference{} van Dyk, D.A., \& Hans, C.M., 2002, in Spatial Cluster Modelling, Eds.\ D.Denison \& A.Lawson, CRC Press: London, p175
\reference{} van Dyk, D.A., \& Kang, H., 2004, Stat.\ Sci, 19, 275
\reference{} Wargelin, B.J., Kashyap, V.L., Drake, J.J., Garc\'{\i}a-Alvarez, D., \& Ratzlaff, P.W., 2006, submitted to ApJ
\reference{} Weisskopf, M.C., Tananbaum, H.D., Van Speybroeck, L.P., \& O'Dell, S.L., 2000, Proc.\ SPIE, 4012, 2 
\reference{} Wichura, M.J., 1989, Technical Report No. 257, Department of Statistics, The University of Chicago
\reference{} Zamorani, G., Henry, J.P., Maccacaro, T., Tananbaum, H.,
Soltan, A., Avni, Y., Liebert, J., Stocke, J., Strittmatter, P.A., Weymann, R.J., Smith, M.G., \& Condon, J.J., 1981, ApJ, 245, 357

\end{references}
\end{document}